\documentclass[longauth]{aa}

\usepackage{graphicx}
\usepackage{natbib}
\usepackage{scalerel}
\usepackage{siunitx}
\usepackage{lastpage}

\usepackage[table]{xcolor}

\usepackage{txfonts}
\usepackage[pdfencoding=auto,psdextra]{hyperref}
\hypersetup{
    colorlinks=true,
    linkcolor=blue,
    filecolor=magenta,      
    urlcolor=blue,
    citecolor=blue
}
\makeatletter
\renewcommand*\aa@pageof{, page \thepage{} of \pageref*{LastPage}}
\makeatother

\usepackage[utf8]{inputenc}

\usepackage[switch, modulo]{lineno}
              
\renewcommand{\linenumbers}[0]{}

\newcommand*{\Gaia}{\textit{Gaia}~DR3\xspace}
\newcommand*{\SCAMP}{\texttt{Scamp}\xspace}
\newcommand*{\SExtractor}{\texttt{SExtractor}\xspace}
\newcommand*{\PSFex}{\texttt{PSFex}\xspace}

\newcommand{\GG}{\ensuremath{G}\xspace}
\newcommand{\BPG}{\ensuremath{{\rm BP}}}
\newcommand{\RPG}{\ensuremath{{\rm RP}}}

\newcommand*{\PE}{\texttt{NIR\_Astrometry}\xspace}
\newcommand*{\OBSID}{\texttt{OBSID}\xspace}

\usepackage{euclid}

\begin{document}

\title{Euclid Data Release (DR1)}
\subtitle{NIR astrometric calibration}

\newcommand{\orcid}[1]{} 
\author{Euclid Collaboration: M.~Radovich\orcid{0000-0002-3585-866X}\thanks{\email{mario.radovich@inaf.it}}\inst{\ref{aff1}}
\and R.~da~Silva\orcid{0000-0003-4788-677X}\inst{\ref{aff2},\ref{aff3}}
\and G.~Polenta\orcid{0000-0003-4067-9196}\inst{\ref{aff3}}
\and A.~Alavi\orcid{0000-0002-8630-6435}\inst{\ref{aff4}}
\and H.~Aussel\orcid{0000-0002-1371-5705}\inst{\ref{aff5}}
\and P.~Awad\orcid{0000-0002-0428-849X}\inst{\ref{aff6}}
\and A.~Bonchi\orcid{0000-0002-2667-5482}\inst{\ref{aff3}}
\and L.~Bramante\inst{\ref{aff7}}
\and S.~Conseil\orcid{0000-0002-3657-4191}\inst{\ref{aff8},\ref{aff9}}
\and M.~Correnti\orcid{0000-0001-6464-3257}\inst{\ref{aff2},\ref{aff3}}
\and L.~V.~Costa\inst{\ref{aff7}}
\and I.~Das\orcid{0009-0007-7088-2044}\inst{\ref{aff4}}
\and F.~Faustini\orcid{0000-0001-6274-5145}\inst{\ref{aff2}}
\and M.~Frailis\orcid{0000-0002-7400-2135}\inst{\ref{aff10}}
\and Y.~Fu\orcid{0000-0002-0759-0504}\inst{\ref{aff6},\ref{aff11}}
\and T.~Gasparetto\orcid{0000-0002-7913-4866}\inst{\ref{aff2}}
\and W.~Gillard\orcid{0000-0003-4744-9748}\inst{\ref{aff12}}
\and S.~Hemmati\orcid{0000-0003-2226-5395}\inst{\ref{aff4}}
\and J.~Jacobson\inst{\ref{aff4}}
\and K.~Jahnke\orcid{0000-0003-3804-2137}\inst{\ref{aff13}}
\and B.~Kubik\orcid{0009-0006-5823-4880}\inst{\ref{aff9}}
\and C.~Macabiau\inst{\ref{aff9}}
\and P.~W.~Morris\orcid{0000-0002-5186-4381}\inst{\ref{aff14}}
\and K.~Paterson\orcid{0000-0001-8340-3486}\inst{\ref{aff13}}
\and M.~Schirmer\orcid{0000-0003-2568-9994}\inst{\ref{aff13}}
\and H.~I.~Teplitz\orcid{0000-0002-7064-5424}\inst{\ref{aff15}}
\and A.~Troja\orcid{0000-0003-0239-4595}\inst{\ref{aff10}}
\and B.~Altieri\orcid{0000-0003-3936-0284}\inst{\ref{aff16}}
\and S.~Andreon\orcid{0000-0002-2041-8784}\inst{\ref{aff17}}
\and N.~Auricchio\orcid{0000-0003-4444-8651}\inst{\ref{aff18}}
\and C.~Baccigalupi\orcid{0000-0002-8211-1630}\inst{\ref{aff19},\ref{aff10},\ref{aff20},\ref{aff21}}
\and M.~Baldi\orcid{0000-0003-4145-1943}\inst{\ref{aff22},\ref{aff18},\ref{aff23}}
\and A.~Balestra\orcid{0000-0002-6967-261X}\inst{\ref{aff1}}
\and P.~Battaglia\orcid{0000-0002-7337-5909}\inst{\ref{aff18}}
\and A.~N.~Belikov\inst{\ref{aff11},\ref{aff24}}
\and A.~Biviano\orcid{0000-0002-0857-0732}\inst{\ref{aff10},\ref{aff19}}
\and M.~Bolzonella\orcid{0000-0003-3278-4607}\inst{\ref{aff18}}
\and E.~Branchini\orcid{0000-0002-0808-6908}\inst{\ref{aff25},\ref{aff26},\ref{aff17}}
\and M.~Brescia\orcid{0000-0001-9506-5680}\inst{\ref{aff27},\ref{aff28}}
\and J.~Brinchmann\orcid{0000-0003-4359-8797}\inst{\ref{aff29},\ref{aff30},\ref{aff31}}
\and S.~Camera\orcid{0000-0003-3399-3574}\inst{\ref{aff32},\ref{aff33},\ref{aff34}}
\and G.~Ca\~nas-Herrera\orcid{0000-0003-2796-2149}\inst{\ref{aff6}}
\and V.~Capobianco\orcid{0000-0002-3309-7692}\inst{\ref{aff34}}
\and C.~Carbone\orcid{0000-0003-0125-3563}\inst{\ref{aff35}}
\and J.~Carretero\orcid{0000-0002-3130-0204}\inst{\ref{aff36},\ref{aff37}}
\and M.~Castellano\orcid{0000-0001-9875-8263}\inst{\ref{aff2}}
\and G.~Castignani\orcid{0000-0001-6831-0687}\inst{\ref{aff18}}
\and S.~Cavuoti\orcid{0000-0002-3787-4196}\inst{\ref{aff28},\ref{aff38}}
\and A.~Cimatti\inst{\ref{aff39}}
\and C.~Colodro-Conde\inst{\ref{aff40}}
\and G.~Congedo\orcid{0000-0003-2508-0046}\inst{\ref{aff41}}
\and C.~J.~Conselice\orcid{0000-0003-1949-7638}\inst{\ref{aff42}}
\and L.~Conversi\orcid{0000-0002-6710-8476}\inst{\ref{aff43},\ref{aff16}}
\and Y.~Copin\orcid{0000-0002-5317-7518}\inst{\ref{aff9}}
\and A.~Costille\inst{\ref{aff44}}
\and H.~M.~Courtois\orcid{0000-0003-0509-1776}\inst{\ref{aff45}}
\and A.~Da~Silva\orcid{0000-0002-6385-1609}\inst{\ref{aff46},\ref{aff47}}
\and H.~Degaudenzi\orcid{0000-0002-5887-6799}\inst{\ref{aff48}}
\and G.~De~Lucia\orcid{0000-0002-6220-9104}\inst{\ref{aff10}}
\and H.~Dole\orcid{0000-0002-9767-3839}\inst{\ref{aff49},\ref{aff50}}
\and F.~Dubath\orcid{0000-0002-6533-2810}\inst{\ref{aff48}}
\and C.~A.~J.~Duncan\orcid{0009-0003-3573-0791}\inst{\ref{aff41}}
\and X.~Dupac\inst{\ref{aff16}}
\and S.~Dusini\orcid{0000-0002-1128-0664}\inst{\ref{aff51}}
\and S.~Escoffier\orcid{0000-0002-2847-7498}\inst{\ref{aff12}}
\and M.~Farina\orcid{0000-0002-3089-7846}\inst{\ref{aff52}}
\and S.~Farrens\orcid{0000-0002-9594-9387}\inst{\ref{aff5}}
\and S.~Ferriol\inst{\ref{aff9}}
\and F.~Finelli\orcid{0000-0002-6694-3269}\inst{\ref{aff18},\ref{aff53}}
\and S.~Fotopoulou\orcid{0000-0002-9686-254X}\inst{\ref{aff54}}
\and N.~Fourmanoit\orcid{0009-0005-6816-6925}\inst{\ref{aff12}}
\and E.~Franceschi\orcid{0000-0002-0585-6591}\inst{\ref{aff18}}
\and M.~Fumana\orcid{0000-0001-6787-5950}\inst{\ref{aff35}}
\and L.~Gabarra\orcid{0000-0002-8486-8856}\inst{\ref{aff55}}
\and S.~Galeotta\orcid{0000-0002-3748-5115}\inst{\ref{aff10}}
\and K.~George\orcid{0000-0002-1734-8455}\inst{\ref{aff56}}
\and B.~Gillis\orcid{0000-0002-4478-1270}\inst{\ref{aff41}}
\and C.~Giocoli\orcid{0000-0002-9590-7961}\inst{\ref{aff18},\ref{aff23}}
\and J.~Gracia-Carpio\orcid{0000-0003-4689-3134}\inst{\ref{aff57}}
\and A.~Grazian\orcid{0000-0002-5688-0663}\inst{\ref{aff1}}
\and F.~Grupp\inst{\ref{aff57},\ref{aff58}}
\and L.~Guzzo\orcid{0000-0001-8264-5192}\inst{\ref{aff59},\ref{aff17},\ref{aff60}}
\and W.~G.~Hartley\orcid{0000-0002-4563-5101}\inst{\ref{aff48}}
\and S.~V.~H.~Haugan\orcid{0000-0001-9648-7260}\inst{\ref{aff61}}
\and H.~Hoekstra\orcid{0000-0002-0641-3231}\inst{\ref{aff6}}
\and W.~Holmes\orcid{0009-0007-8554-4646}\inst{\ref{aff62}}
\and I.~M.~Hook\orcid{0000-0002-2960-978X}\inst{\ref{aff63}}
\and F.~Hormuth\inst{\ref{aff64}}
\and A.~Hornstrup\orcid{0000-0002-3363-0936}\inst{\ref{aff65},\ref{aff66}}
\and M.~Jhabvala\inst{\ref{aff67}}
\and S.~Kermiche\orcid{0000-0002-0302-5735}\inst{\ref{aff12}}
\and A.~Kiessling\orcid{0000-0002-2590-1273}\inst{\ref{aff62}}
\and R.~Kohley\inst{\ref{aff16}}
\and M.~K\"ummel\orcid{0000-0003-2791-2117}\inst{\ref{aff58}}
\and M.~Kunz\orcid{0000-0002-3052-7394}\inst{\ref{aff68}}
\and H.~Kurki-Suonio\orcid{0000-0002-4618-3063}\inst{\ref{aff69},\ref{aff70}}
\and A.~M.~C.~Le~Brun\orcid{0000-0002-0936-4594}\inst{\ref{aff71}}
\and D.~Le~Mignant\orcid{0000-0002-5339-5515}\inst{\ref{aff44}}
\and S.~Ligori\orcid{0000-0003-4172-4606}\inst{\ref{aff34}}
\and P.~B.~Lilje\orcid{0000-0003-4324-7794}\inst{\ref{aff61}}
\and V.~Lindholm\orcid{0000-0003-2317-5471}\inst{\ref{aff69},\ref{aff70}}
\and I.~Lloro\orcid{0000-0001-5966-1434}\inst{\ref{aff72}}
\and M.~Magliocchetti\orcid{0000-0001-9158-4838}\inst{\ref{aff52}}
\and G.~Mainetti\orcid{0000-0003-2384-2377}\inst{\ref{aff73}}
\and O.~Mansutti\orcid{0000-0001-5758-4658}\inst{\ref{aff10}}
\and S.~Marcin\inst{\ref{aff74}}
\and O.~Marggraf\orcid{0000-0001-7242-3852}\inst{\ref{aff75}}
\and M.~Martinelli\orcid{0000-0002-6943-7732}\inst{\ref{aff2},\ref{aff76}}
\and N.~Martinet\orcid{0000-0003-2786-7790}\inst{\ref{aff44}}
\and F.~Marulli\orcid{0000-0002-8850-0303}\inst{\ref{aff77},\ref{aff18},\ref{aff23}}
\and R.~J.~Massey\orcid{0000-0002-6085-3780}\inst{\ref{aff78}}
\and N.~Mauri\orcid{0000-0001-8196-1548}\inst{\ref{aff39},\ref{aff23}}
\and H.~J.~McCracken\orcid{0000-0002-9489-7765}\inst{\ref{aff79}}
\and C.~J.~R.~McPartland\orcid{0000-0003-0639-025X}\inst{\ref{aff66},\ref{aff80}}
\and E.~Medinaceli\orcid{0000-0002-4040-7783}\inst{\ref{aff18}}
\and S.~Mei\orcid{0000-0002-2849-559X}\inst{\ref{aff81},\ref{aff82}}
\and M.~Melchior\inst{\ref{aff83}}
\and M.~Meneghetti\orcid{0000-0003-1225-7084}\inst{\ref{aff18},\ref{aff23}}
\and E.~Merlin\orcid{0000-0001-6870-8900}\inst{\ref{aff1}}
\and G.~Meylan\orcid{0000-0001-6503-0209}\inst{\ref{aff84}}
\and P.~Monaco\orcid{0000-0003-2083-7564}\inst{\ref{aff85},\ref{aff10},\ref{aff20},\ref{aff19}}
\and A.~Mora\orcid{0000-0002-1922-8529}\inst{\ref{aff86}}
\and M.~Moresco\orcid{0000-0002-7616-7136}\inst{\ref{aff77},\ref{aff18}}
\and C.~Moretti\orcid{0000-0003-3314-8936}\inst{\ref{aff10},\ref{aff19},\ref{aff20}}
\and L.~Moscardini\orcid{0000-0002-3473-6716}\inst{\ref{aff77},\ref{aff18},\ref{aff23}}
\and R.~Nakajima\orcid{0009-0009-1213-7040}\inst{\ref{aff75}}
\and C.~Neissner\orcid{0000-0001-8524-4968}\inst{\ref{aff87},\ref{aff37}}
\and R.~C.~Nichol\orcid{0000-0003-0939-6518}\inst{\ref{aff88}}
\and S.-M.~Niemi\orcid{0009-0005-0247-0086}\inst{\ref{aff89}}
\and C.~Padilla\orcid{0000-0001-7951-0166}\inst{\ref{aff87}}
\and S.~Paltani\orcid{0000-0002-8108-9179}\inst{\ref{aff48}}
\and F.~Pasian\orcid{0000-0002-4869-3227}\inst{\ref{aff10}}
\and W.~J.~Percival\orcid{0000-0002-0644-5727}\inst{\ref{aff90},\ref{aff91},\ref{aff92}}
\and V.~Pettorino\orcid{0000-0002-4203-9320}\inst{\ref{aff89}}
\and A.~Pezzotta\orcid{0000-0003-0726-2268}\inst{\ref{aff17}}
\and M.~Poncet\inst{\ref{aff93}}
\and L.~A.~Popa\inst{\ref{aff94}}
\and F.~Raison\orcid{0000-0002-7819-6918}\inst{\ref{aff57}}
\and A.~Refregier\inst{\ref{aff95}}
\and A.~Renzi\orcid{0000-0001-9856-1970}\inst{\ref{aff96},\ref{aff51},\ref{aff18}}
\and J.~Rhodes\orcid{0000-0002-4485-8549}\inst{\ref{aff62}}
\and G.~Riccio\inst{\ref{aff28}}
\and I.~Risso\orcid{0000-0003-2525-7761}\inst{\ref{aff25},\ref{aff26},\ref{aff17}}
\and H.-W.~Rix\orcid{0000-0003-4996-9069}\inst{\ref{aff13}}
\and F.~Rizzo\orcid{0000-0002-9407-585X}\inst{\ref{aff10}}
\and E.~Romelli\orcid{0000-0003-3069-9222}\inst{\ref{aff10}}
\and C.~Rosset\orcid{0000-0003-0286-2192}\inst{\ref{aff81}}
\and B.~Rusholme\orcid{0000-0001-7648-4142}\inst{\ref{aff4}}
\and R.~Saglia\orcid{0000-0003-0378-7032}\inst{\ref{aff58},\ref{aff57}}
\and Z.~Sakr\orcid{0000-0002-4823-3757}\inst{\ref{aff97},\ref{aff98},\ref{aff99}}
\and A.~G.~S\'anchez\orcid{0000-0003-1198-831X}\inst{\ref{aff57}}
\and D.~Sapone\orcid{0000-0001-7089-4503}\inst{\ref{aff100}}
\and B.~Sartoris\orcid{0000-0003-1337-5269}\inst{\ref{aff101}}
\and P.~Schneider\orcid{0000-0001-8561-2679}\inst{\ref{aff75}}
\and A.~Secroun\orcid{0000-0003-0505-3710}\inst{\ref{aff12}}
\and E.~Sihvola\orcid{0000-0003-1804-7715}\inst{\ref{aff69},\ref{aff70}}
\and C.~Sirignano\orcid{0000-0002-0995-7146}\inst{\ref{aff96},\ref{aff51}}
\and G.~Sirri\orcid{0000-0003-2626-2853}\inst{\ref{aff23}}
\and L.~Stanco\orcid{0000-0002-9706-5104}\inst{\ref{aff51}}
\and P.~Tallada-Cresp\'{i}\orcid{0000-0002-1336-8328}\inst{\ref{aff36},\ref{aff37}}
\and A.~N.~Taylor\inst{\ref{aff41}}
\and I.~Tereno\orcid{0000-0002-4537-6218}\inst{\ref{aff46},\ref{aff47}}
\and N.~Tessore\orcid{0000-0002-9696-7931}\inst{\ref{aff102}}
\and S.~Toft\orcid{0000-0003-3631-7176}\inst{\ref{aff103},\ref{aff80}}
\and R.~Toledo-Moreo\orcid{0000-0002-2997-4859}\inst{\ref{aff104},\ref{aff105}}
\and F.~Torradeflot\orcid{0000-0003-1160-1517}\inst{\ref{aff37},\ref{aff36}}
\and A.~Tsyganov\inst{\ref{aff106}}
\and I.~Tutusaus\orcid{0000-0002-3199-0399}\inst{\ref{aff107},\ref{aff108},\ref{aff98}}
\and J.~Valiviita\orcid{0000-0001-6225-3693}\inst{\ref{aff69},\ref{aff70}}
\and T.~Vassallo\orcid{0000-0001-6512-6358}\inst{\ref{aff10},\ref{aff56}}
\and A.~Veropalumbo\orcid{0000-0003-2387-1194}\inst{\ref{aff17},\ref{aff26},\ref{aff25}}
\and Y.~Wang\orcid{0000-0002-4749-2984}\inst{\ref{aff4}}
\and J.~Weller\orcid{0000-0002-8282-2010}\inst{\ref{aff58},\ref{aff57}}
\and A.~Zacchei\orcid{0000-0003-0396-1192}\inst{\ref{aff10},\ref{aff19}}
\and G.~Zamorani\orcid{0000-0002-2318-301X}\inst{\ref{aff18}}
\and F.~M.~Zerbi\orcid{0000-0002-9996-973X}\inst{\ref{aff17}}
\and E.~Zucca\orcid{0000-0002-5845-8132}\inst{\ref{aff18}}
\and I.~T.~Andika\orcid{0000-0001-6102-9526}\inst{\ref{aff58}}
\and M.~Ballardini\orcid{0000-0003-4481-3559}\inst{\ref{aff109},\ref{aff110},\ref{aff18}}
\and P.~Bergamini\orcid{0000-0003-1383-9414}\inst{\ref{aff18}}
\and S.~Borgani\orcid{0000-0001-6151-6439}\inst{\ref{aff85},\ref{aff19},\ref{aff10},\ref{aff20},\ref{aff111}}
\and E.~Bozzo\orcid{0000-0002-8201-1525}\inst{\ref{aff48}}
\and C.~Burigana\orcid{0000-0002-3005-5796}\inst{\ref{aff112},\ref{aff53}}
\and R.~Cabanac\orcid{0000-0001-6679-2600}\inst{\ref{aff98}}
\and M.~Calabrese\orcid{0000-0002-2637-2422}\inst{\ref{aff113},\ref{aff35}}
\and A.~Cappi\inst{\ref{aff101},\ref{aff18}}
\and F.~Caro\orcid{0009-0003-1053-0507}\inst{\ref{aff2}}
\and T.~Castro\orcid{0000-0002-6292-3228}\inst{\ref{aff114},\ref{aff10},\ref{aff19}}
\and D.~Di~Ferdinando\inst{\ref{aff23}}
\and J.~A.~Escartin~Vigo\inst{\ref{aff57}}
\and G.~Fabbian\orcid{0000-0002-3255-4695}\inst{\ref{aff49}}
\and J.~Garc\'ia-Bellido\orcid{0000-0002-9370-8360}\inst{\ref{aff97}}
\and C.~M.~Gutierrez\orcid{0000-0001-7854-783X}\inst{\ref{aff40},\ref{aff115}}
\and A.~Hall\orcid{0000-0002-3139-8651}\inst{\ref{aff41}}
\and Y.~Kang\orcid{0009-0000-8588-7250}\inst{\ref{aff116}}
\and A.~Loureiro\orcid{0000-0002-4371-0876}\inst{\ref{aff117},\ref{aff118}}
\and J.~Macias-Perez\orcid{0000-0002-5385-2763}\inst{\ref{aff119}}
\and R.~Maoli\orcid{0000-0002-6065-3025}\inst{\ref{aff120},\ref{aff2}}
\and R.~B.~Metcalf\orcid{0000-0003-3167-2574}\inst{\ref{aff77},\ref{aff18}}
\and M.~P\"ontinen\orcid{0000-0001-5442-2530}\inst{\ref{aff69}}
\and E.~Sarpa\orcid{0000-0002-1256-655X}\inst{\ref{aff10}}
\and V.~Scottez\orcid{0009-0008-3864-940X}\inst{\ref{aff121},\ref{aff122}}
\and M.~Sereno\orcid{0000-0003-0302-0325}\inst{\ref{aff18},\ref{aff23}}
\and K.~Tanidis\orcid{0000-0001-9843-5130}\inst{\ref{aff123}}
\and M.~Tenti\orcid{0000-0002-4254-5901}\inst{\ref{aff23}}
\and M.~Tucci\inst{\ref{aff48}}
\and M.~Viel\orcid{0000-0002-2642-5707}\inst{\ref{aff19},\ref{aff10},\ref{aff21},\ref{aff20},\ref{aff111}}
\and M.~Wiesmann\orcid{0009-0000-8199-5860}\inst{\ref{aff61}}
\and J.~A.~Acevedo~Barroso\orcid{0000-0002-9654-1711}\inst{\ref{aff62}}
\and Y.~Akrami\orcid{0000-0002-2407-7956}\inst{\ref{aff97},\ref{aff124}}
\and G.~Angora\orcid{0000-0002-0316-6562}\inst{\ref{aff28},\ref{aff109}}
\and S.~Anselmi\orcid{0000-0002-3579-9583}\inst{\ref{aff51},\ref{aff96},\ref{aff125}}
\and M.~Archidiacono\orcid{0000-0003-4952-9012}\inst{\ref{aff59},\ref{aff60}}
\and G.~Aric\`o\orcid{0000-0002-2802-2928}\inst{\ref{aff23}}
\and F.~Atrio-Barandela\orcid{0000-0002-2130-2513}\inst{\ref{aff126}}
\and M.~Aubert\orcid{0009-0002-7667-8814}\inst{\ref{aff9}}
\and M.~Baes\orcid{0000-0002-3930-2757}\inst{\ref{aff127}}
\and L.~Bazzanini\orcid{0000-0003-0727-0137}\inst{\ref{aff109},\ref{aff18}}
\and J.~R.~Bermejo-Climent\inst{\ref{aff128},\ref{aff129}}
\and D.~Bertacca\orcid{0000-0002-2490-7139}\inst{\ref{aff96},\ref{aff1},\ref{aff51}}
\and M.~Bethermin\orcid{0000-0002-3915-2015}\inst{\ref{aff130}}
\and F.~Beutler\orcid{0000-0003-0467-5438}\inst{\ref{aff41}}
\and A.~Blanchard\orcid{0000-0001-8555-9003}\inst{\ref{aff98}}
\and F.~Bouch\`e\orcid{0000-0002-4663-1786}\inst{\ref{aff131},\ref{aff38}}
\and M.~L.~Brown\orcid{0000-0002-0370-8077}\inst{\ref{aff42}}
\and S.~Bruton\orcid{0000-0002-6503-5218}\inst{\ref{aff14}}
\and A.~Calabro\orcid{0000-0003-2536-1614}\inst{\ref{aff2}}
\and B.~Camacho~Quevedo\orcid{0000-0002-8789-4232}\inst{\ref{aff19},\ref{aff21},\ref{aff10}}
\and P.~Carrilho\orcid{0000-0003-1339-0194}\inst{\ref{aff132}}
\and C.~S.~Carvalho\inst{\ref{aff133}}
\and Y.~Charles\orcid{0009-0000-3636-834X}\inst{\ref{aff44}}
\and F.~Cogato\orcid{0000-0003-4632-6113}\inst{\ref{aff18}}
\and T.~E.~Collett\orcid{0000-0001-5564-3140}\inst{\ref{aff134}}
\and A.~R.~Cooray\orcid{0000-0002-3892-0190}\inst{\ref{aff135}}
\and P.~Corcho-Caballero\orcid{0000-0001-6327-7080}\inst{\ref{aff11}}
\and B.~Csizi\orcid{0000-0003-3227-6581}\inst{\ref{aff116}}
\and O.~Cucciati\orcid{0000-0002-9336-7551}\inst{\ref{aff18}}
\and T.~de~Boer\orcid{0000-0001-5486-2747}\inst{\ref{aff136}}
\and G.~Desprez\orcid{0000-0001-8325-1742}\inst{\ref{aff11}}
\and A.~D\'iaz-S\'anchez\orcid{0000-0003-0748-4768}\inst{\ref{aff137}}
\and S.~Di~Domizio\orcid{0000-0003-2863-5895}\inst{\ref{aff25},\ref{aff26}}
\and J.~M.~Diego\orcid{0000-0001-9065-3926}\inst{\ref{aff138}}
\and V.~Duret\orcid{0009-0009-0383-4960}\inst{\ref{aff12}}
\and A.~Enia\orcid{0000-0002-0200-2857}\inst{\ref{aff139},\ref{aff18}}
\and Y.~Fang\orcid{0000-0002-0334-6950}\inst{\ref{aff58}}
\and A.~Farina\orcid{0009-0000-3420-929X}\inst{\ref{aff17},\ref{aff26}}
\and A.~M.~N.~Ferguson\orcid{0000-0001-7934-1278}\inst{\ref{aff41}}
\and A.~Finoguenov\orcid{0000-0002-4606-5403}\inst{\ref{aff69}}
\and A.~Franco\orcid{0000-0002-4761-366X}\inst{\ref{aff140},\ref{aff141},\ref{aff142}}
\and A.~Fumagalli\orcid{0009-0004-0300-2535}\inst{\ref{aff10}}
\and K.~Ganga\orcid{0000-0001-8159-8208}\inst{\ref{aff81}}
\and R.~Gavazzi\orcid{0000-0002-5540-6935}\inst{\ref{aff44},\ref{aff79}}
\and E.~Gaztanaga\orcid{0000-0001-9632-0815}\inst{\ref{aff107},\ref{aff108},\ref{aff134}}
\and Z.~Ghaffari\orcid{0000-0002-6467-8078}\inst{\ref{aff10},\ref{aff19}}
\and F.~Giacomini\orcid{0000-0002-3129-2814}\inst{\ref{aff23}}
\and F.~Gianotti\orcid{0000-0003-4666-119X}\inst{\ref{aff18}}
\and G.~Gozaliasl\orcid{0000-0002-0236-919X}\inst{\ref{aff143},\ref{aff69}}
\and A.~Gruppuso\orcid{0000-0001-9272-5292}\inst{\ref{aff18},\ref{aff23}}
\and M.~Guidi\orcid{0000-0001-9408-1101}\inst{\ref{aff22},\ref{aff18}}
\and C.~Hern\'andez-Monteagudo\orcid{0000-0001-5471-9166}\inst{\ref{aff115},\ref{aff40}}
\and T.~Heuberger\orcid{0009-0009-3474-8467}\inst{\ref{aff74}}
\and H.~Hildebrandt\orcid{0000-0002-9814-3338}\inst{\ref{aff144}}
\and J.~Hjorth\orcid{0000-0002-4571-2306}\inst{\ref{aff145}}
\and M.~Jauzac\orcid{0000-0003-1974-8732}\inst{\ref{aff146},\ref{aff78},\ref{aff147},\ref{aff148},\ref{aff98}}
\and J.~J.~E.~Kajava\orcid{0000-0002-3010-8333}\inst{\ref{aff149},\ref{aff150},\ref{aff151}}
\and V.~Kansal\orcid{0000-0002-4008-6078}\inst{\ref{aff152},\ref{aff153}}
\and J.~Kim\orcid{0000-0003-2776-2761}\inst{\ref{aff55}}
\and C.~C.~Kirkpatrick\inst{\ref{aff154}}
\and A.~Kov\'acs\orcid{0000-0002-5825-579X}\inst{\ref{aff129},\ref{aff128}}
\and I.~Kova{\v{c}}i{\'{c}}\orcid{0000-0001-6751-3263}\inst{\ref{aff127}}
\and K.~Koyama\orcid{0000-0001-6727-6915}\inst{\ref{aff134}}
\and S.~Kruk\orcid{0000-0001-8010-8879}\inst{\ref{aff16}}
\and M.~C.~Lam\orcid{0000-0002-9347-2298}\inst{\ref{aff41}}
\and M.~Lattanzi\orcid{0000-0003-1059-2532}\inst{\ref{aff110}}
\and S.~Lee\orcid{0000-0002-8289-740X}\inst{\ref{aff155},\ref{aff62}}
\and L.~Legrand\orcid{0000-0003-0610-5252}\inst{\ref{aff156},\ref{aff49}}
\and M.~D.~Lepinzan\orcid{0000-0003-1287-9801}\inst{\ref{aff10}}
\and G.~Leroy\orcid{0009-0004-2523-4425}\inst{\ref{aff146},\ref{aff78}}
\and J.~Lesgourgues\orcid{0000-0001-7627-353X}\inst{\ref{aff157}}
\and T.~I.~Liaudat\orcid{0000-0002-9104-314X}\inst{\ref{aff158}}
\and L.~Linke\orcid{0000-0002-2622-8113}\inst{\ref{aff116}}
\and S.~J.~Liu\orcid{0000-0001-7680-2139}\inst{\ref{aff52}}
\and G.~Maggio\orcid{0000-0003-4020-4836}\inst{\ref{aff10}}
\and F.~Mannucci\orcid{0000-0002-4803-2381}\inst{\ref{aff159}}
\and F.~R.~Marleau\orcid{0000-0002-1442-2947}\inst{\ref{aff116}}
\and C.~J.~A.~P.~Martins\orcid{0000-0002-4886-9261}\inst{\ref{aff160},\ref{aff29}}
\and M.~Miluzio\inst{\ref{aff16},\ref{aff161}}
\and G.~Morgante\inst{\ref{aff18}}
\and S.~Nadathur\orcid{0000-0001-9070-3102}\inst{\ref{aff134}}
\and K.~Naidoo\orcid{0000-0002-9182-1802}\inst{\ref{aff134},\ref{aff13}}
\and P.~Natoli\orcid{0000-0003-0126-9100}\inst{\ref{aff109},\ref{aff110}}
\and A.~Navarro-Alsina\orcid{0000-0002-3173-2592}\inst{\ref{aff75}}
\and S.~Nesseris\orcid{0000-0002-0567-0324}\inst{\ref{aff97}}
\and F.~Oppizzi\orcid{0000-0003-3904-8370}\inst{\ref{aff26},\ref{aff51},\ref{aff96}}
\and F.~Pace\orcid{0000-0001-8039-0480}\inst{\ref{aff32},\ref{aff33},\ref{aff34},\ref{aff47}}
\and D.~Paoletti\orcid{0000-0003-4761-6147}\inst{\ref{aff18},\ref{aff53}}
\and F.~Passalacqua\orcid{0000-0002-8606-4093}\inst{\ref{aff51}}
\and L.~Patrizii\inst{\ref{aff23}}
\and C.~Pattison\orcid{0000-0003-3272-2617}\inst{\ref{aff134}}
\and A.~Pisani\orcid{0000-0002-6146-4437}\inst{\ref{aff12}}
\and L.~Pizzuti\orcid{0000-0001-5654-7580}\inst{\ref{aff162}}
\and D.~Potter\orcid{0000-0002-0757-5195}\inst{\ref{aff163}}
\and M.~Poulain\orcid{0000-0002-7664-4510}\inst{\ref{aff164}}
\and G.~W.~Pratt\inst{\ref{aff5}}
\and S.~Quai\orcid{0000-0002-0449-8163}\inst{\ref{aff77},\ref{aff18}}
\and G.~Rodighiero\orcid{0000-0002-9415-2296}\inst{\ref{aff96},\ref{aff1}}
\and W.~Roster\orcid{0000-0002-9149-6528}\inst{\ref{aff57}}
\and M.~Sahl\'en\orcid{0000-0003-0973-4804}\inst{\ref{aff165}}
\and D.~B.~Sanders\orcid{0000-0002-1233-9998}\inst{\ref{aff136}}
\and A.~Schneider\orcid{0000-0001-7055-8104}\inst{\ref{aff163}}
\and M.~Schultheis\inst{\ref{aff101}}
\and D.~Sciotti\orcid{0009-0008-4519-2620}\inst{\ref{aff2},\ref{aff76}}
\and E.~Sellentin\orcid{0009-0002-2655-3458}\inst{\ref{aff166},\ref{aff6}}
\and S.~Serjeant\orcid{0000-0002-0517-7943}\inst{\ref{aff167}}
\and L.~C.~Smith\orcid{0000-0002-3259-2771}\inst{\ref{aff168}}
\and J.~G.~Sorce\orcid{0000-0002-2307-2432}\inst{\ref{aff169},\ref{aff49}}
\and I.~Szapudi\orcid{0000-0003-2274-0301}\inst{\ref{aff136}}
\and M.~Talia\orcid{0000-0003-4352-2063}\inst{\ref{aff77},\ref{aff18}}
\and F.~Tarsitano\orcid{0000-0002-5919-0238}\inst{\ref{aff170},\ref{aff95}}
\and G.~Testera\orcid{0000-0003-2970-766X}\inst{\ref{aff26}}
\and R.~Teyssier\orcid{0000-0001-7689-0933}\inst{\ref{aff171}}
\and S.~Tosi\orcid{0000-0002-7275-9193}\inst{\ref{aff25},\ref{aff17},\ref{aff26}}
\and C.~Uhlemann\orcid{0000-0001-7831-1579}\inst{\ref{aff172},\ref{aff173}}
\and C.~Valieri\inst{\ref{aff23}}
\and A.~Venhola\orcid{0000-0001-6071-4564}\inst{\ref{aff164}}
\and D.~Vergani\orcid{0000-0003-0898-2216}\inst{\ref{aff18}}
\and P.~Vielzeuf\orcid{0000-0003-2035-9339}\inst{\ref{aff12}}
\and E.~Vilenius\orcid{0000-0002-6184-7681}\inst{\ref{aff102}}
\and S.~Vinciguerra\orcid{0009-0005-4018-3184}\inst{\ref{aff44}}
\and M.~von~Wietersheim-Kramsta\orcid{0000-0003-4986-5091}\inst{\ref{aff78},\ref{aff146}}
\and N.~A.~Walton\orcid{0000-0003-3983-8778}\inst{\ref{aff168}}
\and A.~H.~Wright\orcid{0000-0001-7363-7932}\inst{\ref{aff144}}
\and D.~Scott\orcid{0000-0002-6878-9840}\inst{\ref{aff174}}}
										   
\institute{INAF-Osservatorio Astronomico di Padova, Via dell'Osservatorio 5, 35122 Padova, Italy\label{aff1}
\and
INAF-Osservatorio Astronomico di Roma, Via Frascati 33, 00078 Monteporzio Catone, Italy\label{aff2}
\and
Space Science Data Center, Italian Space Agency, via del Politecnico snc, 00133 Roma, Italy\label{aff3}
\and
Caltech/IPAC, 1200 E. California Blvd., Pasadena, CA 91125, USA\label{aff4}
\and
Universit\'e Paris-Saclay, Universit\'e Paris Cit\'e, CEA, CNRS, AIM, 91191, Gif-sur-Yvette, France\label{aff5}
\and
Leiden Observatory, Leiden University, Einsteinweg 55, 2333 CC Leiden, The Netherlands\label{aff6}
\and
Aerospace Logistics Technology Engineering Company, Corso Marche 79, 10146 Torino, Italy\label{aff7}
\and
Centre de Recherche Astrophysique de Lyon, UMR5574, CNRS, Universit\'e Claude Bernard Lyon 1, ENS de Lyon, 69230, Saint-Genis-Laval, France\label{aff8}
\and
Universit\'e Lyon 1, CNRS/IN2P3, IP2I Lyon, Villeurbanne, France\label{aff9}
\and
INAF-Osservatorio Astronomico di Trieste, Via G. B. Tiepolo 11, 34143 Trieste, Italy\label{aff10}
\and
Kapteyn Astronomical Institute, University of Groningen, PO Box 800, 9700 AV Groningen, The Netherlands\label{aff11}
\and
Aix-Marseille Universit\'e, CNRS/IN2P3, CPPM, Marseille, France\label{aff12}
\and
Max-Planck-Institut f\"ur Astronomie, K\"onigstuhl 17, 69117 Heidelberg, Germany\label{aff13}
\and
California Institute of Technology, 1200 E California Blvd, Pasadena, CA 91125, USA\label{aff14}
\and
Infrared Processing and Analysis Center, California Institute of Technology, Pasadena, CA 91125, USA\label{aff15}
\and
ESAC/ESA, Camino Bajo del Castillo, s/n., Urb. Villafranca del Castillo, 28692 Villanueva de la Ca\~nada, Madrid, Spain\label{aff16}
\and
INAF-Osservatorio Astronomico di Brera, Via Brera 28, 20122 Milano, Italy\label{aff17}
\and
INAF-Osservatorio di Astrofisica e Scienza dello Spazio di Bologna, Via Piero Gobetti 93/3, 40129 Bologna, Italy\label{aff18}
\and
IFPU, Institute for Fundamental Physics of the Universe, via Beirut 2, 34151 Trieste, Italy\label{aff19}
\and
INFN, Sezione di Trieste, Via Valerio 2, 34127 Trieste TS, Italy\label{aff20}
\and
SISSA, International School for Advanced Studies, Via Bonomea 265, 34136 Trieste TS, Italy\label{aff21}
\and
Dipartimento di Fisica e Astronomia, Universit\`a di Bologna, Via Gobetti 93/2, 40129 Bologna, Italy\label{aff22}
\and
INFN-Sezione di Bologna, Viale Berti Pichat 6/2, 40127 Bologna, Italy\label{aff23}
\and
ATG Europe BV, Huygensstraat 34, 2201 DK Noordwijk, The Netherlands\label{aff24}
\and
Dipartimento di Fisica, Universit\`a di Genova, Via Dodecaneso 33, 16146, Genova, Italy\label{aff25}
\and
INFN-Sezione di Genova, Via Dodecaneso 33, 16146, Genova, Italy\label{aff26}
\and
Department of Physics "E. Pancini", University Federico II, Via Cinthia 6, 80126, Napoli, Italy\label{aff27}
\and
INAF-Osservatorio Astronomico di Capodimonte, Via Moiariello 16, 80131 Napoli, Italy\label{aff28}
\and
Instituto de Astrof\'isica e Ci\^encias do Espa\c{c}o, Universidade do Porto, CAUP, Rua das Estrelas, PT4150-762 Porto, Portugal\label{aff29}
\and
Faculdade de Ci\^encias da Universidade do Porto, Rua do Campo de Alegre, 4150-007 Porto, Portugal\label{aff30}
\and
European Southern Observatory, Karl-Schwarzschild-Str.~2, 85748 Garching, Germany\label{aff31}
\and
Dipartimento di Fisica, Universit\`a degli Studi di Torino, Via P. Giuria 1, 10125 Torino, Italy\label{aff32}
\and
INFN-Sezione di Torino, Via P. Giuria 1, 10125 Torino, Italy\label{aff33}
\and
INAF-Osservatorio Astrofisico di Torino, Via Osservatorio 20, 10025 Pino Torinese (TO), Italy\label{aff34}
\and
INAF-IASF Milano, Via Alfonso Corti 12, 20133 Milano, Italy\label{aff35}
\and
Centro de Investigaciones Energ\'eticas, Medioambientales y Tecnol\'ogicas (CIEMAT), Avenida Complutense 40, 28040 Madrid, Spain\label{aff36}
\and
Port d'Informaci\'{o} Cient\'{i}fica, Campus UAB, C. Albareda s/n, 08193 Bellaterra (Barcelona), Spain\label{aff37}
\and
INFN -- Sezione di Napoli, Via Cinthia 6, 80126, Napoli, Italy\label{aff38}
\and
Dipartimento di Fisica e Astronomia "Augusto Righi" - Alma Mater Studiorum Universit\`a di Bologna, Viale Berti Pichat 6/2, 40127 Bologna, Italy\label{aff39}
\and
Instituto de Astrof\'{\i}sica de Canarias, E-38205 La Laguna, Tenerife, Spain\label{aff40}
\and
Institute for Astronomy, University of Edinburgh, Royal Observatory, Blackford Hill, Edinburgh EH9 3HJ, UK\label{aff41}
\and
Jodrell Bank Centre for Astrophysics, Department of Physics and Astronomy, University of Manchester, Oxford Road, Manchester M13 9PL, UK\label{aff42}
\and
European Space Agency/ESRIN, Largo Galileo Galilei 1, 00044 Frascati, Roma, Italy\label{aff43}
\and
Aix-Marseille Universit\'e, CNRS, CNES, LAM, Marseille, France\label{aff44}
\and
Universit\'e Lyon 1, CNRS/IN2P3, IUF, IP2I Lyon, Villeurbanne, France \label{aff45}
\and
Departamento de F\'isica, Faculdade de Ci\^encias, Universidade de Lisboa, Edif\'icio C8, Campo Grande, PT1749-016 Lisboa, Portugal\label{aff46}
\and
Instituto de Astrof\'isica e Ci\^encias do Espa\c{c}o, Faculdade de Ci\^encias, Universidade de Lisboa, Campo Grande, 1749-016 Lisboa, Portugal\label{aff47}
\and
Department of Astronomy, University of Geneva, ch. d'Ecogia 16, 1290 Versoix, Switzerland\label{aff48}
\and
Universit\'e Paris-Saclay, CNRS, Institut d'astrophysique spatiale, 91405, Orsay, France\label{aff49}
\and
Institut universitaire de France (IUF), 1 rue Descartes, 75231 PARIS CEDEX 05, France\label{aff50}
\and
INFN-Padova, Via Marzolo 8, 35131 Padova, Italy\label{aff51}
\and
INAF-Istituto di Astrofisica e Planetologia Spaziali, via del Fosso del Cavaliere, 100, 00100 Roma, Italy\label{aff52}
\and
INFN-Bologna, Via Irnerio 46, 40126 Bologna, Italy\label{aff53}
\and
School of Physics, HH Wills Physics Laboratory, University of Bristol, Tyndall Avenue, Bristol, BS8 1TL, UK\label{aff54}
\and
Department of Physics, University of Oxford, Keble Road, Oxford OX1 3RH, UK\label{aff55}
\and
University Observatory, LMU Faculty of Physics, Scheinerstr.~1, 81679 Munich, Germany\label{aff56}
\and
Max Planck Institute for Extraterrestrial Physics, Giessenbachstr. 1, 85748 Garching, Germany\label{aff57}
\and
Universit\"ats-Sternwarte M\"unchen, Fakult\"at f\"ur Physik, Ludwig-Maximilians-Universit\"at M\"unchen, Scheinerstr.~1, 81679 M\"unchen, Germany\label{aff58}
\and
Dipartimento di Fisica "Aldo Pontremoli", Universit\`a degli Studi di Milano, Via Celoria 16, 20133 Milano, Italy\label{aff59}
\and
INFN-Sezione di Milano, Via Celoria 16, 20133 Milano, Italy\label{aff60}
\and
Institute of Theoretical Astrophysics, University of Oslo, P.O. Box 1029 Blindern, 0315 Oslo, Norway\label{aff61}
\and
Jet Propulsion Laboratory, California Institute of Technology, 4800 Oak Grove Drive, Pasadena, CA, 91109, USA\label{aff62}
\and
School of Physics and Astronomy, Lancaster University, Lancaster, LA1 4YB, UK\label{aff63}
\and
Felix Hormuth Engineering, Goethestr. 17, 69181 Leimen, Germany\label{aff64}
\and
Technical University of Denmark, Elektrovej 327, 2800 Kgs. Lyngby, Denmark\label{aff65}
\and
Cosmic Dawn Center (DAWN), Denmark\label{aff66}
\and
NASA Goddard Space Flight Center, Greenbelt, MD 20771, USA\label{aff67}
\and
Universit\'e de Gen\`eve, D\'epartement de Physique Th\'eorique and Centre for Astroparticle Physics, 24 quai Ernest-Ansermet, CH-1211 Gen\`eve 4, Switzerland\label{aff68}
\and
Department of Physics, P.O. Box 64, University of Helsinki, 00014 Helsinki, Finland\label{aff69}
\and
Helsinki Institute of Physics, Gustaf H{\"a}llstr{\"o}min katu 2, University of Helsinki, 00014 Helsinki, Finland\label{aff70}
\and
Laboratoire d'etude de l'Univers et des phenomenes eXtremes, Observatoire de Paris, Universit\'e PSL, Sorbonne Universit\'e, CNRS, 92190 Meudon, France\label{aff71}
\and
SKAO, Jodrell Bank, Lower Withington, Macclesfield SK11 9FT, UK\label{aff72}
\and
Centre de Calcul de l'IN2P3/CNRS, 21 avenue Pierre de Coubertin 69627 Villeurbanne Cedex, France\label{aff73}
\and
University of Applied Sciences and Arts of Northwestern Switzerland, School of Computer Science, 5210 Windisch, Switzerland\label{aff74}
\and
Universit\"at Bonn, Argelander-Institut f\"ur Astronomie, Auf dem H\"ugel 71, 53121 Bonn, Germany\label{aff75}
\and
INFN-Sezione di Roma, Piazzale Aldo Moro, 2 - c/o Dipartimento di Fisica, Edificio G. Marconi, 00185 Roma, Italy\label{aff76}
\and
Dipartimento di Fisica e Astronomia "Augusto Righi" - Alma Mater Studiorum Universit\`a di Bologna, via Piero Gobetti 93/2, 40129 Bologna, Italy\label{aff77}
\and
Department of Physics, Institute for Computational Cosmology, Durham University, South Road, Durham, DH1 3LE, UK\label{aff78}
\and
Institut d'Astrophysique de Paris, UMR 7095, CNRS, and Sorbonne Universit\'e, 98 bis boulevard Arago, 75014 Paris, France\label{aff79}
\and
Niels Bohr Institute, University of Copenhagen, Jagtvej 128, 2200 Copenhagen, Denmark\label{aff80}
\and
Universit\'e Paris Cit\'e, CNRS, Astroparticule et Cosmologie, 75013 Paris, France\label{aff81}
\and
CNRS-UCB International Research Laboratory, Centre Pierre Bin\'etruy, IRL2007, CPB-IN2P3, Berkeley, USA\label{aff82}
\and
University of Applied Sciences and Arts of Northwestern Switzerland, School of Engineering, 5210 Windisch, Switzerland\label{aff83}
\and
Institute of Physics, Laboratory of Astrophysics, Ecole Polytechnique F\'ed\'erale de Lausanne (EPFL), Observatoire de Sauverny, 1290 Versoix, Switzerland\label{aff84}
\and
Dipartimento di Fisica - Sezione di Astronomia, Universit\`a di Trieste, Via Tiepolo 11, 34131 Trieste, Italy\label{aff85}
\and
Telespazio UK S.L. for European Space Agency (ESA), Camino bajo del Castillo, s/n, Urbanizacion Villafranca del Castillo, Villanueva de la Ca\~nada, 28692 Madrid, Spain\label{aff86}
\and
Institut de F\'{i}sica d'Altes Energies (IFAE), The Barcelona Institute of Science and Technology, Campus UAB, 08193 Bellaterra (Barcelona), Spain\label{aff87}
\and
School of Mathematics and Physics, University of Surrey, Guildford, Surrey, GU2 7XH, UK\label{aff88}
\and
European Space Agency/ESTEC, Keplerlaan 1, 2201 AZ Noordwijk, The Netherlands\label{aff89}
\and
Waterloo Centre for Astrophysics, University of Waterloo, Waterloo, Ontario N2L 3G1, Canada\label{aff90}
\and
Department of Physics and Astronomy, University of Waterloo, Waterloo, Ontario N2L 3G1, Canada\label{aff91}
\and
Perimeter Institute for Theoretical Physics, Waterloo, Ontario N2L 2Y5, Canada\label{aff92}
\and
Centre National d'Etudes Spatiales -- Centre spatial de Toulouse, 18 avenue Edouard Belin, 31401 Toulouse Cedex 9, France\label{aff93}
\and
Institute of Space Science, Str. Atomistilor, nr. 409 M\u{a}gurele, Ilfov, 077125, Romania\label{aff94}
\and
Institute for Particle Physics and Astrophysics, Dept. of Physics, ETH Zurich, Wolfgang-Pauli-Strasse 27, 8093 Zurich, Switzerland\label{aff95}
\and
Dipartimento di Fisica e Astronomia "G. Galilei", Universit\`a di Padova, Via Marzolo 8, 35131 Padova, Italy\label{aff96}
\and
Instituto de F\'isica Te\'orica UAM-CSIC, Campus de Cantoblanco, 28049 Madrid, Spain\label{aff97}
\and
Institut de Recherche en Astrophysique et Plan\'etologie (IRAP), Universit\'e de Toulouse, CNRS, UPS, CNES, 14 Av. Edouard Belin, 31400 Toulouse, France\label{aff98}
\and
Universit\'e St Joseph; Faculty of Sciences, Beirut, Lebanon\label{aff99}
\and
Departamento de F\'isica, FCFM, Universidad de Chile, Blanco Encalada 2008, Santiago, Chile\label{aff100}
\and
Universit\'e C\^{o}te d'Azur, Observatoire de la C\^{o}te d'Azur, CNRS, Laboratoire Lagrange, Bd de l'Observatoire, CS 34229, 06304 Nice cedex 4, France\label{aff101}
\and
Mullard Space Science Laboratory, University College London, Holmbury St Mary, Dorking, Surrey RH5 6NT, UK\label{aff102}
\and
Cosmic Dawn Center (DAWN)\label{aff103}
\and
Universidad Polit\'ecnica de Cartagena, Departamento de Electr\'onica y Tecnolog\'ia de Computadoras,  Plaza del Hospital 1, 30202 Cartagena, Spain\label{aff104}
\and
European University of Technology EUt+, European Union\label{aff105}
\and
Centre for Information Technology, University of Groningen, P.O. Box 11044, 9700 CA Groningen, The Netherlands\label{aff106}
\and
Institute of Space Sciences (ICE, CSIC), Campus UAB, Carrer de Can Magrans, s/n, 08193 Barcelona, Spain\label{aff107}
\and
Institut d'Estudis Espacials de Catalunya (IEEC),  Edifici RDIT, Campus UPC, 08860 Castelldefels, Barcelona, Spain\label{aff108}
\and
Dipartimento di Fisica e Scienze della Terra, Universit\`a degli Studi di Ferrara, Via Giuseppe Saragat 1, 44122 Ferrara, Italy\label{aff109}
\and
Istituto Nazionale di Fisica Nucleare, Sezione di Ferrara, Via Giuseppe Saragat 1, 44122 Ferrara, Italy\label{aff110}
\and
ICSC - Centro Nazionale di Ricerca in High Performance Computing, Big Data e Quantum Computing, Via Magnanelli 2, Bologna, Italy\label{aff111}
\and
INAF, Istituto di Radioastronomia, Via Piero Gobetti 101, 40129 Bologna, Italy\label{aff112}
\and
Astronomical Observatory of the Autonomous Region of the Aosta Valley (OAVdA), Loc. Lignan 39, I-11020, Nus (Aosta Valley), Italy\label{aff113}
\and
Department of Mathematical Physics, Institute of Physics, University of S\~ao Paulo, R. do Mat\~ao 1371, 05508-090, S\~ao Paulo, SP, Brazil\label{aff114}
\and
Universidad de La Laguna, Dpto. Astrof\'\i sica, E-38206 La Laguna, Tenerife, Spain\label{aff115}
\and
Universit\"at Innsbruck, Institut f\"ur Astro- und Teilchenphysik, Technikerstr. 25/8, 6020 Innsbruck, Austria\label{aff116}
\and
Oskar Klein Centre for Cosmoparticle Physics, Department of Physics, Stockholm University, Stockholm, SE-106 91, Sweden\label{aff117}
\and
Astrophysics Group, Blackett Laboratory, Imperial College London, London SW7 2AZ, UK\label{aff118}
\and
Univ. Grenoble Alpes, CNRS, Grenoble INP, LPSC-IN2P3, 53, Avenue des Martyrs, 38000 Grenoble, France\label{aff119}
\and
Dipartimento di Fisica, Sapienza Universit\`a di Roma, Piazzale Aldo Moro 2, 00185 Roma, Italy\label{aff120}
\and
Institut d'Astrophysique de Paris, 98bis Boulevard Arago, 75014, Paris, France\label{aff121}
\and
ICL, Junia, Universit\'e Catholique de Lille, LITL, 59000 Lille, France\label{aff122}
\and
Center for Astrophysics and Cosmology, University of Nova Gorica, Nova Gorica, Slovenia\label{aff123}
\and
CERCA/ISO, Department of Physics, Case Western Reserve University, 10900 Euclid Avenue, Cleveland, OH 44106, USA\label{aff124}
\and
Laboratoire Univers et Th\'eorie, Observatoire de Paris, Universit\'e PSL, Universit\'e Paris Cit\'e, CNRS, 92190 Meudon, France\label{aff125}
\and
Departamento de F{\'\i}sica Fundamental. Universidad de Salamanca. Plaza de la Merced s/n. 37008 Salamanca, Spain\label{aff126}
\and
Universiteit Gent, Department of Physics and Astronomy, Proeftuinstraat 86 N3, 9000 Ghent, Belgium
\label{aff127}
\and
Konkoly Observatory, HUN-REN CSFK, MTA Centre of Excellence, Budapest, Konkoly Thege Mikl\'os {\'u}t 15-17. H-1121, Hungary\label{aff128}
\and
MTA-CSFK Lend\"ulet Large-Scale Structure Research Group, Konkoly-Thege Mikl\'os \'ut 15-17, H-1121 Budapest, Hungary\label{aff129}
\and
Universit\'e de Strasbourg, CNRS, Observatoire astronomique de Strasbourg, UMR 7550, 67000 Strasbourg, France\label{aff130}
\and
Scuola Superiore Meridionale, Via Mezzocannone 4, 80138, Napoli, Italy\label{aff131}
\and
Department of Physics, Astronomy and Mathematics, University of Hertfordshire, College Lane, Hatfield AL10 9AB, UK\label{aff132}
\and
Instituto de Astrof\'isica e Ci\^encias do Espa\c{c}o, Faculdade de Ci\^encias, Universidade de Lisboa, Tapada da Ajuda, 1349-018 Lisboa, Portugal\label{aff133}
\and
Institute of Cosmology and Gravitation, University of Portsmouth, Portsmouth PO1 3FX, UK\label{aff134}
\and
Department of Physics \& Astronomy, University of California Irvine, Irvine CA 92697, USA\label{aff135}
\and
Institute for Astronomy, University of Hawaii, 2680 Woodlawn Drive, Honolulu, HI 96822, USA\label{aff136}
\and
Departamento F\'isica Aplicada, Universidad Polit\'ecnica de Cartagena, Campus Muralla del Mar, 30202 Cartagena, Murcia, Spain\label{aff137}
\and
Instituto de F\'isica de Cantabria, Edificio Juan Jord\'a, Avenida de los Castros, 39005 Santander, Spain\label{aff138}
\and
CEA Saclay, DFR/IRFU, Service d'Astrophysique, Bat. 709, 91191 Gif-sur-Yvette, France\label{aff139}
\and
Department of Mathematics and Physics E. De Giorgi, University of Salento, Via per Arnesano, CP-I93, 73100, Lecce, Italy\label{aff140}
\and
INFN, Sezione di Lecce, Via per Arnesano, CP-193, 73100, Lecce, Italy\label{aff141}
\and
INAF-Sezione di Lecce, c/o Dipartimento Matematica e Fisica, Via per Arnesano, 73100, Lecce, Italy\label{aff142}
\and
Department of Computer Science, Aalto University, PO Box 15400, Espoo, FI-00 076, Finland\label{aff143}
\and
Ruhr University Bochum, Faculty of Physics and Astronomy, Astronomical Institute (AIRUB), German Centre for Cosmological Lensing (GCCL), 44780 Bochum, Germany\label{aff144}
\and
DARK, Niels Bohr Institute, University of Copenhagen, Jagtvej 155, 2200 Copenhagen, Denmark\label{aff145}
\and
Department of Physics, Centre for Extragalactic Astronomy, Durham University, South Road, Durham, DH1 3LE, UK\label{aff146}
\and
Astrophysics Research Centre, University of KwaZulu-Natal, Westville Campus, Durban 4041, South Africa\label{aff147}
\and
School of Mathematics, Statistics \& Computer Science, University of KwaZulu-Natal, Westville Campus, Durban 4041, South Africa\label{aff148}
\and
Department of Physics and Astronomy, Vesilinnantie 5, University of Turku, 20014 Turku, Finland\label{aff149}
\and
Finnish Centre for Astronomy with ESO (FINCA), Quantum, Vesilinnantie 5, University of Turku, 20014 Turku, Finland\label{aff150}
\and
Serco for European Space Agency (ESA), Camino bajo del Castillo, s/n, Urbanizacion Villafranca del Castillo, Villanueva de la Ca\~nada, 28692 Madrid, Spain\label{aff151}
\and
ARC Centre of Excellence for Dark Matter Particle Physics, Melbourne, Australia\label{aff152}
\and
Centre for Astrophysics \& Supercomputing, Swinburne University of Technology,  Hawthorn, Victoria 3122, Australia\label{aff153}
\and
Department of Physics and Helsinki Institute of Physics, Gustaf H\"allstr\"omin katu 2, University of Helsinki, 00014 Helsinki, Finland\label{aff154}
\and
Ohio University, Physics \& Astronomy Department,1 Ohio University, Athens, OH 45701, USA\label{aff155}
\and
Brazilian Center for Research in Physics (CBPF), Dr. Xavier Sigaud st. 150, zip 22290-180, Rio de Janeiro, RJ, Brazil\label{aff156}
\and
Institute for Theoretical Particle Physics and Cosmology (TTK), RWTH Aachen University, 52056 Aachen, Germany\label{aff157}
\and
IRFU, CEA, Universit\'e Paris-Saclay 91191 Gif-sur-Yvette Cedex, France\label{aff158}
\and
INAF-Osservatorio Astrofisico di Arcetri, Largo E. Fermi 5, 50125, Firenze, Italy\label{aff159}
\and
Centro de Astrof\'{\i}sica da Universidade do Porto, Rua das Estrelas, 4150-762 Porto, Portugal\label{aff160}
\and
HE Space for European Space Agency (ESA), Camino bajo del Castillo, s/n, Urbanizacion Villafranca del Castillo, Villanueva de la Ca\~nada, 28692 Madrid, Spain\label{aff161}
\and
Dipartimento di Fisica ``G. Occhialini", Universit\`a degli Studi di Milano Bicocca, Piazza della Scienza 3, 20126 Milano, Italy\label{aff162}
\and
Department of Astrophysics, University of Zurich, Winterthurerstrasse 190, 8057 Zurich, Switzerland\label{aff163}
\and
Space physics and astronomy research unit, University of Oulu, Pentti Kaiteran katu 1, FI-90014 Oulu, Finland\label{aff164}
\and
Theoretical astrophysics, Department of Physics and Astronomy, Uppsala University, Box 524, 751 20 Uppsala, Sweden\label{aff165}
\and
Mathematical Institute, University of Leiden, Einsteinweg 55, 2333 CA Leiden, The Netherlands\label{aff166}
\and
School of Physical Sciences, The Open University, Milton Keynes, MK7 6AA, UK\label{aff167}
\and
Institute of Astronomy, University of Cambridge, Madingley Road, Cambridge CB3 0HA, UK\label{aff168}
\and
Univ. Lille, CNRS, Centrale Lille, UMR 9189 CRIStAL, 59000 Lille, France\label{aff169}
\and
Kobayashi-Maskawa Institute for the Origin of Particles and the Universe, Nagoya University, Chikusa-ku, Nagoya, 464-8602, Japan\label{aff170}
\and
Department of Astrophysical Sciences, Peyton Hall, Princeton University, Princeton, NJ 08544, USA\label{aff171}
\and
Fakult\"at f\"ur Physik, Universit\"at Bielefeld, Postfach 100131, 33501 Bielefeld, Germany\label{aff172}
\and
School of Mathematics, Statistics and Physics, Newcastle University, Herschel Building, Newcastle-upon-Tyne, NE1 7RU, UK\label{aff173}
\and
Department of Physics and Astronomy, University of British Columbia, Vancouver, BC V6T 1Z1, Canada\label{aff174}}    
 \date\today

\abstract{
This paper presents the methodology and results of the astrometric calibration applied to the \Euclid Near-Infrared (NIR) images in the Euclid Data Release 1, focusing on the techniques used to model and correct instrumental distortions, as well as the assessment of calibration quality. This work details the calibration pipeline, including quality control tests, and the evaluation of astrometric residuals against the \Gaia catalogue and internal consistency metrics based on data encompassing more than one year of \Euclid operation: the achieved accuracy is  30 mas for 97\% of the total area, thus exceeding the requirements for Data Release 1. For the remaining 3\%, accuracy declines due to a persistence correction issue: a method to correct the issue is presented. Furthermore, we discuss the stability of the solution over time, its dependence on source properties and instrument configuration, and outline the improvements planned for future \Euclid data releases to enhance the astrometric calibration accuracy and efficiency. 
    }

\keywords{Cosmology: observations -- Surveys -- Astrometry -- Techniques: image processing  -- Space vehicles: instruments -- Instrumentation: detectors}

\titlerunning{NIR astrometric calibration}
   
   \maketitle 
%
%
%
%
   
\section{\label{sc:Intro}Introduction}
Accurate astrometric calibration is essential for the success of the \Euclid\ mission \citep{EuclidSkyOverview}, which enables the precise mapping of celestial sources across its detectors. The astrometric calibration process ensures the reliable alignment of observed images with external reference catalogues, such as \Gaia \citep{GaiaDR3_2023A&A...674A...1G}, and maintains internal consistency among multiple dithered images.
\Euclid's  Near-Infrared Spectrometer and Photometer \citep[NISP:][]{EuclidSkyNISP} features three imaging channels: \YE, \JE, and \HE.
An array of 16 HAWAII-2RG (H2RG) detectors, each  with 2048x2048 pixels and a plate scale of \ang{;;0.3} pixel$^{-1}$, provides a field of view of 0.57~$\mathrm{deg}^2$. 
We recall 
that within \Euclid, a dithering sequence operating mode is termed a single observation, and the individual exposures (dithers) of each observation are archived under an Observation ID (\OBSID hereafter). 
In the wide survey, each \OBSID consists of four dithers; the self-calibration field consists of 76 dithers \citep{EuclidSkyOverview, Q1-TP003}.
In DR1 \citep{DR1cite}, NISP calibrated images and associated catalogues are available for both single dithers and stacked products \citep{DR1-TP003}.   

For the initial data releases (DR1 and DR2), the  astrometric calibration requirement is set as a root mean square (RMS) of 200\,mas for both RA and Dec, compared to the external reference (absolute astrometry), and 100\,mas for the positions of the same sources observed within a dither sequence (relative astrometry). The ultimate aim of the final release (DR3) is to achieve an RMS of 30\,mas on each axis for both relative and absolute astrometry,  corresponding to 0.1 pixels.
An additional requirement pertains to geometric distortion, which must be calibrated to within 0.015\% over scales of 3\,arcmin.
This paper focuses on the methods used to model and correct instrumental distortions in images  from the NISP photometric channel  (NISP-P) and assesses the calibration quality achieved in the Euclid DR1, covering an area of 
1900~$\mathrm{deg}^2$ \citep{Scaramella-EP1}. 
The paper is organized as follows. Section~\ref{sc:Pipeline} summarises the astrometric calibration module of the NIR processing function (NIR-PF) pipeline and its quality control tests. Section~\ref{sc:Analysis} presents the results obtained in the DR1, evaluating astrometric residuals against external catalogues and internal consistency metrics based on more than 1 year of \Euclid operations.
This analysis shows that  the astrometric calibration meets the requirements on the RMS and on the geometric distortion accuracy, described in Sect.~\ref{sc:stability}.
The stability of the astrometric solution over time and its dependence on source properties and instrument configuration are also discussed in Sect.~\ref{sc:stability}. Section~\ref{sc:improvements} outlines the enhancements already implemented for Euclid Data Release 2, as well as additional improvements  that may   further refine the calibration accuracy and efficiency in later data releases.
To support the use of the calibrated data products, for each parameter
described in the text we also give the corresponding keyword in the
\texttt{NIR\_AstrometricCalibration} module of the NIR-PF (\PE\ hereafter).

\section{\label{sc:Pipeline}Astrometric calibration of  NISP images in Euclid DR1}
In the NIR-PF, \PE relies on the \SCAMP \footnote{The NIR-PF adopts a rengineered version of \SCAMP, named as \texttt{EL\_ScampPP}.} software  \citep{scamp}  to compute the distortion 
and update the associated keywords in the headers of the {\tt FITS} frames.  This is done  for each dither, because the combined effect of the tilt in the NISP optics \citep{Schirmer-EP18} and an offset of the filter centre from the optical axis, due to uncertainty in filter positioning, leads to variations in the distortion pattern that prevent us assuming a constant distortion across different exposures. 

\begin{figure} 
\centering 
\includegraphics[width=1.0\linewidth]{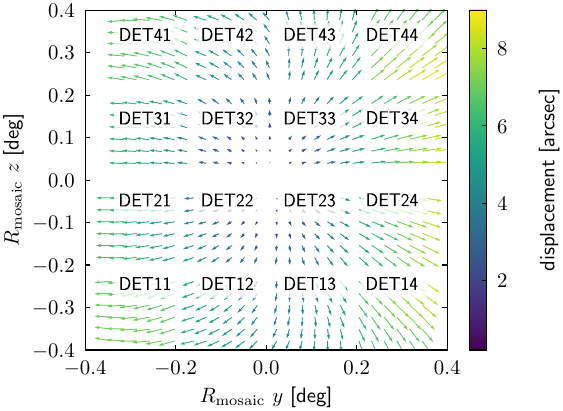} \caption{Distortion map showing the displacement  (arcsec) on the focal plane in the \texttt{R-mosaic} coordinate system \citep{EuclidSkyNISP}, as derived from a self-calibration observation. } \label{fig:distortion}
\end{figure}

\SCAMP utilises the {\tt PV} formalism \citep[see e.g.][]{Calabretta2002, Schupe2012}, which calculates the distortion for each detector by mapping pixel coordinates ($x$, $y$) from that detector to intermediate world coordinates ($\xi$, $\eta$). The distortion is applied to the coordinates projected onto the focal plane as
\begin{equation}
    \begin{aligned}
        X &= C_{11}(x - p_{\rm{x}}) + C_{12}(y - p_{\rm{y}}) \, , \\[4pt]
        Y &= C_{21}(x - p_{\rm{x}}) + C_{22}(y - p_{\rm{y}}) \, , \\[8pt]
        \xi &= \sum_{m+n \le N} a_{mn} \, X^m Y^n, \\[4pt]
        \eta &= \sum_{m+n \le N} b_{mn} \, X^m Y^n.
    \end{aligned}
    \label{eqn:PV}
\end{equation}
The relevant quantities are defined as follows: $p_{\mathrm{x}}$ and $p_{\mathrm{y}}$ are the coordinates of the reference pixel, corresponding to the {\tt CRPIX}  keywords in the {\tt FITS} files; $C_{ij}$ are the elements of the linear transformation matrix that accounts for pixel scale and  rotation, stored in the  {\tt CD} keywords; and $a_{mn}$ and $b_{mn}$ are for each axis the coefficients of the polynomial with order $N$ describing the distortion, corresponding to the {\tt PV} keywords.
In  \PE, the distortion (Fig.~\ref{fig:distortion}) is represented  by a polynomial with $N=3$.
This approach involves working in sky coordinates using a tangent-plane projection. In cases such as \Euclid, where the detectors are not aligned with the celestial coordinates, the polynomial coefficients are dependent  on the telescope rotation.
In contrast, the {\tt SIP} formalism \citep{Schupe2005} describes the distortion directly in the focal plane without relying on the 
tangent-plane projection. It uses polynomial distortion coefficients $A_{ij}$ and $B_{ij}$ (often encoded as {\tt SIP} keywords) to model the optical and instrumental distortions that affect the pixel coordinates  on the detector and transform them into intermediate world coordinates:
\begin{equation}
    \begin{aligned}
        u &= x - p_{{\rm{x}}}, &\qquad v &= y - p_{\rm{y}} \, , \\[8pt]
        U &= u + \sum_{i+j \le N} A_{ij} \, u^i v^j, &\qquad V &= v + \sum_{i+j \le N} B_{ij} \, u^i v^j \, , \\[8pt]
        \xi &= C_{11} U + C_{12} V \, , &\qquad \eta &= C_{21} U + C_{22} V \, .
    \end{aligned}
    \label{eqn:SIP}
\end{equation}
In this case, because the distortion is expressed purely in focal plane terms, it does not depend on the telescope rotation. 

In DR1, the adopted external reference is the  \Gaia catalogue \citep{Gaia2016, Gaia_DR3}, which is also the reference catalogue adopted for the astrometric calibration of the VIS channel. The positions from the VIS catalogue are not utilised for the following reasons. (1) Although there is a significant overlap between the VIS and NISP coverage in each exposure, it is not complete; therefore, incorporating additional adjacent VIS fields becomes necessary, but these fields are not guaranteed to be observed. (2) 
Point sources brighter than \IE=18.5 are saturated in VIS normal exposures \citep{Q1-TP002}; as a consequence,  a significant fraction of  bright  sources that are not saturated in NISP\footnote{In NISP, the saturation limit for bright sources is approximately 16.0 AB mag \citep{EuclidSkyNISP}.} and therefore available for astrometric calibration with \Gaia as reference,  would be saturated in VIS and thus not  usable with VIS as reference.

 The \Gaia celestial coordinates (epoch J2016.0) are  propagated to the Euclid dither epoch through the application of proper motions. The final Euclid calibrated positions are in the barycentric ICRS.

 \SCAMP processes the catalogues generated by \SExtractor \citep{sextractor} in combination with \PSFex \citep{psfex}, optimising the precision of source centres through point-spread function (PSF) fitting. 
 The implementation of PSF fitting in the NIR-PF is discussed in detail in \citet{DR1-TP005}.
 Within \SCAMP, the quality maps produced by the NIR-PF \citep{Q1-TP003} enable the exclusion of sources that are either saturated or compromised by issues like cosmic rays and persistence. The sources used for astrometric calibration have \Gaia magnitudes, \GG,  in the 15 to 21 range, corresponding to approximately 16 to 22.5 in the NIR bands.
 
 The robustness of astrometric calibration can be enhanced by supplying \SCAMP with an initial estimate of the distortion (presolution). This approach is particularly advantageous for the detectors at the edges  of the field of view (Fig.~\ref{fig:distortion}), where distortion is more pronounced, thereby complicating the alignment with the external reference catalogue. In the current implementation, the presolution primarily serves as an initial input to \SCAMP, functioning as a preliminary estimate or starting point rather than delivering a highly precise outcome. In this implementation, because the presolution is primarily used to guide the matching of sources before further processing in \SCAMP, the demand for accuracy is relatively low (around 1\arcsec).
Although the distortion pattern is expected to vary among different exposures, deriving the presolution once is sufficient for the intended purpose of the current implementation. This derivation is based on the observations of the self-calibration field.
The presolution initially contains distortion terms expressed in the \SCAMP\ {\tt PV} format, along with the values encoded in the {\tt CD} and {\tt CRPIX} keywords. To facilitate its application to observations with varying telescope rotation angles, the presolution is initially transformed into the {\tt SIP} format. Subsequently, it is converted back to the PV format, with modifications made to the {\tt CD} matrix terms to accommodate the instrument rotation. Analytical expressions derived using the {\tt sympy} package in {\tt Python} are applied for all conversions between the {\tt PV} and {\tt SIP} formats.
At this point, there may be an offset in the celestial coordinates of the tangent-point, defined by the {\tt CRVAL} keyword in the images.   Although \SCAMP\ corrects this offset before calculating the distortion, it is ineffective in situations such as a Fine Guidance Sensor (FGS) failure, which results in elongated source images, or when cosmic ray showers impact a significant number of  pixels, necessitating the exclusion of the exposures regardless. Consequently, \PE initially employs \SCAMP  exclusively to adjust the {\tt CRVAL} values, followed by an initial alignment utilising the \Gaia catalogue through the presolution. Subsequently, a series of tests is conducted for each dither, to identify those for which it is not possible to derive a reliable astrometric solution.
\begin{itemize}    
\item   If the median ellipticity of the matched sources ({\tt ASTRELL} keyword)  exceeds a threshold of $\epsilon_{\rm th}=0.3$ or is more than twice   the median across all dithers, it is highly likely that the dither is compromised owing to a failure in the FGS.  
\item If a detector records an abnormally low number of detections ($< 5$) or if the fraction of matched sources is too low compared to those in the external catalogue ($<10\%$), it becomes impossible to reliably estimate the distortion.
\end{itemize}
The threshold values mentioned above are selected based on an analysis of a subset of dithers with known issues and may be refined in future data releases. For the dithers that successfully pass these evaluations, \SCAMP\ is  executed once more to calculate the total distortion. 
One-third of the dithers identified as invalid in the NIR-PF \citep{DR1-TP003} are marked as such through this step. Section~\ref{sc:analysis_dithers} presents an overview of  the astrometric calibration results for dithers that pass the validity tests. 

\section{Analysis of astrometric residuals \label{sc:Analysis}}
We present a summary of the astrometric residuals generated by  \PE, both in relation to the \Gaia catalogue and internally among the dithers. To accomplish this, we select 4000 \OBSID from the Euclid Wide Survey in the Science Archive System (SAS) for each NISP filter, covering an area of  1950~$\mathrm{deg}^2$ after removing overlaps (Sect.~\ref{sc:overlaps}), over more than 1 year of operation (from February 2024 to May 2025).

To assess the quality of the astrometric calibration, we use the following
quantities derived from the positional differences between the $N$
detections and their matched reference sources,
$\Delta\alpha^{*}_{i}=(\alpha_i-\alpha_{{\rm ref},i})\cos\delta_{{\rm ref},i}$
in right ascension (RA) and $\Delta\delta_i=\delta_i-\delta_{{\rm ref},i}$ in
declination (Dec). Quantities in RA marked with an asterisk indicate that they are multiplied by $\cos\delta$ to measure
angular distances on the sky.

\begin{description}\item[\textbf{Average offset.}] The mean positional difference, which indicates the calibration bias,  is given by 
\begin{equation}
\langle \Delta \alpha^{*} \rangle =  \frac{1}{N} \sum_{i=1}^{N} \Delta\alpha^{*}_{i} \, , \quad\langle \Delta \delta \rangle =  \frac{1}{N} \sum_{i=1}^{N} \Delta\delta_i \, .
\label{eqn:avgoff_def}
\end{equation}

\item[\textbf{Standard deviation ($\sigma$).}] This measures the spread of the positional offsets around the mean, thereby indicating the calibration precision,
\begin{equation}
 \sigma_{\alpha^{*}}^2 = \frac{1}{N} \sum_{i=1}^{N} \left(\Delta \alpha^{*}_{i} - \langle \Delta \alpha^{*} \rangle \right)^2 \, ,  \quad \sigma_\delta^2 = \frac{1}{N} \sum_{i=1}^{N} \left(\Delta \delta_i - \langle \Delta\delta \rangle \right)^2 \, .
  \label{eqn:sigma_def}
 \end{equation}
\item[\textbf{RMS offset.}] This is the square root of the mean squared positional differences, representing the total calibration accuracy (bias + precision),
\begin{equation}
\mathrm{RMS}_{\alpha*}^2    = \frac{1}{N} \sum_{i=1}^{N} (\Delta\alpha^{*}_{i})^2 \, ,     \quad    \mathrm{RMS}_\delta^2    = \frac{1}{N} \sum_{i=1}^{N} (\Delta\delta_i)^2 \, .
 \label{eqn:rms_def}
\end{equation}
Where the RMS or $\sigma$ represents the sum over both coordinates, it is  expressed as   $\mathrm{RMS}^2=\left(\mathrm{RMS}_{\alpha*}^2 + \mathrm{RMS}_\delta^2\right)/2$\,.
\end{description}

In \PE, the average offset relative to the reference catalogue is produced as an output and stored in the headers of the calibrated images using the keywords {\tt ASTROFF1} and {\tt ASTROFF2}, corresponding to the RA and Dec coordinates, respectively. The keywords  {\tt ASTIRMS1}, {\tt ASTIRMS2}
represent the standard deviation of the residuals for sources   matched across different dithers;
{\tt ASTRRMS1}, and {\tt ASTRRMS2} refer to  those compared to the reference catalogue. They are computed both for single detectors and the entire dither. Given that the average offset in each detector or dither is negligible, the standard deviation matches the RMS of the deviations. 
During the validation process, dithers are flagged if any value of {\tt ASTIRMS1}, {\tt ASTIRMS2}, {\tt ASTRRMS1}, or {\tt ASTRRMS2} computed throughout the entire dither exceed 50\,mas or if {\tt ASTRIRMS1} or {\tt ASTIRMS2} exceed 50\,mas in any detector.

\begin{figure*}
    \centering
    \includegraphics[width=1.0\linewidth]{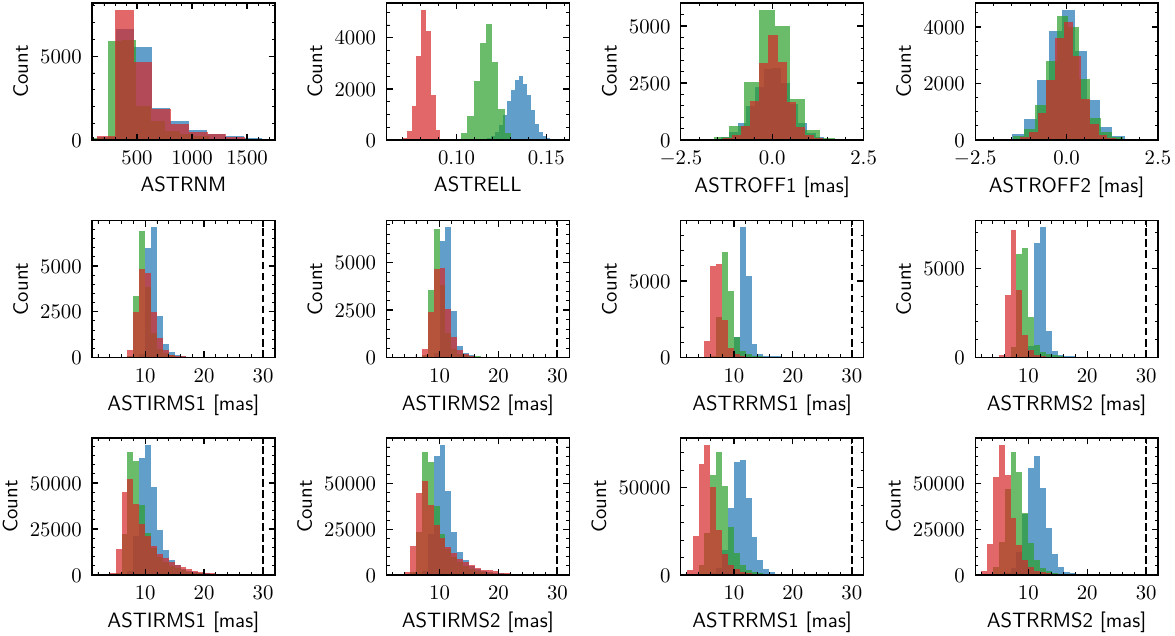}
    \caption{
    Histograms illustrating the following  for each of the \YE (blue), \JE (green), and \HE (red) filters: the number of matched sources available for astrometric calibration ({\tt ASTRNM}); the median ellipticity ({\tt ASTRELL}); the average offsets ({\tt ASTROFF1}, {\tt ASTROFF2}); the standard deviations of residuals for sources matched across dithers ({\tt ASTIRMS1}, {\tt ASTIRMS2}), and compared to the reference catalogue {\tt (ASTRRMS1}, {\tt ASTRRMS2}).
    The top two rows pertain to dithers, while the bottom row corresponds to values within individual detectors. The dashed line represents the final requirement for the RMS, set at 30\,mas.}
    \label{fig:stats_dithers}
\end{figure*}

\subsection {Astrometric residuals in dithers\label{sc:analysis_dithers}}

Figure~\ref{fig:stats_dithers} illustrates the distribution of average ellipticities, the number of matches with \Gaia sources, and the residuals (offsets and standard deviations) in both the dithers (upper and middle rows) and individual detectors (bottom row). This confirms that, as anticipated in Sect.~\ref{sc:Analysis}, the average offset to the reference catalogue is indeed negligible, being within 2\,mas; the astrometric calibration accuracy is approximately 10\,mas and remains below 20\,mas for 99\% of the dithers, in all bands. The RMS decreases from \YE\ to \JE\ and \HE, consistent with the stronger undersampling of the diffraction-limited PSF at shorter wavelengths \citep{EuclidSkyNISP, DR1-TP005}. 
For individual detectors, mostly DET13 and DET14, a small fraction exhibits a standard deviation exceeding 60\,mas and potentially reaching up to 150\,mas. This fraction is at most 0.15\% for \HE and is even lower in the other bands. As discussed in more detail in Sects.~\ref{sc:persistence} and \ref{sc:persistence_corr}, the problem was identified as stemming from an excessively stringent exclusion of detections linked to the persistence flagging.

Figure~\ref{fig:stats_dithers} also confirms that, among the dithers that passed the initial selection, the average ellipticity is less than 0.16. In future data releases, a threshold lower than the current value  ($\epsilon_{\rm th}=0.3$) could be chosen to identify dithers affected by the FGS issue.

\begin{figure*}
    \centering
    \includegraphics[width=.9\linewidth]
    {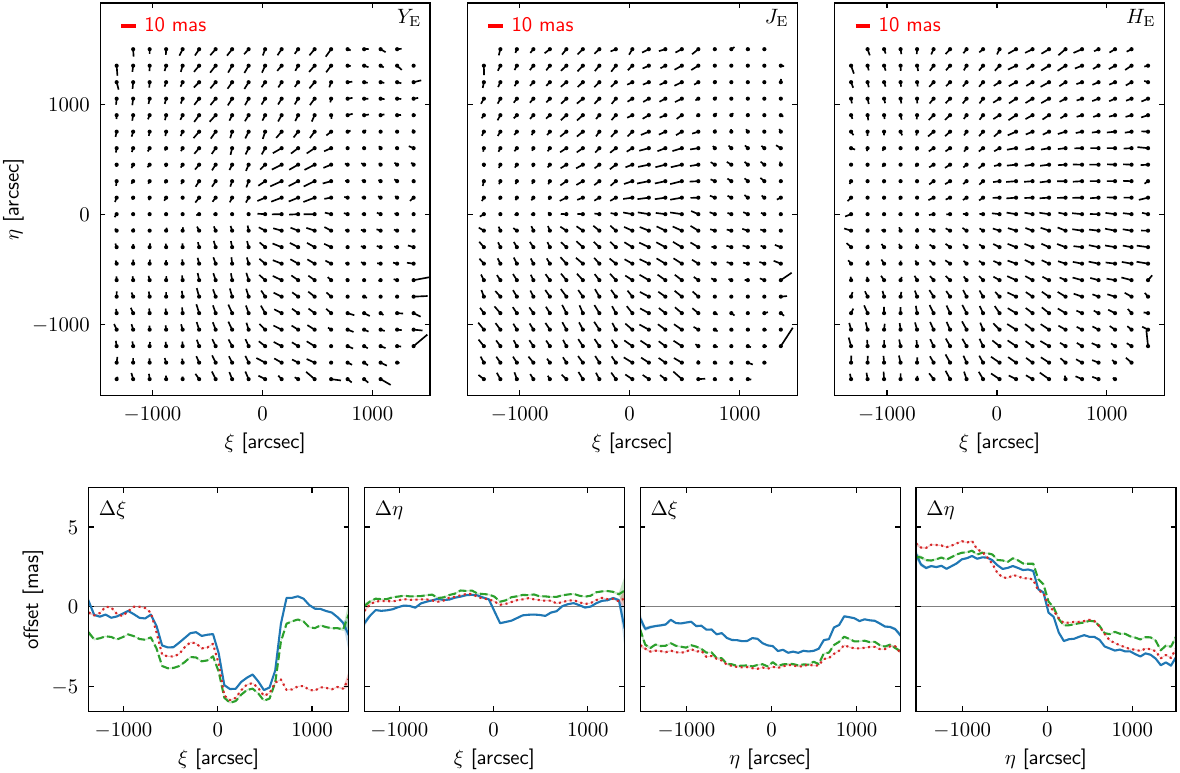}
\caption{Clipped mean astrometric residuals  in the instrument frame, obtained combining the measurements from the stacked catalogues. Top: whiskers with components
$(\langle\Delta\xi\rangle,\langle\Delta\eta\rangle)$ derived in  $2\farcm5\times2\farcm5$ cells.
Bottom: $\Delta\xi$ and $\Delta\eta$ as a function of $\xi$ and $\eta$ for
\YE (blue), \JE (green), and \HE (red).}

    \label{fig:fp_offsets}
\end{figure*}

\begin{figure*}
    \centering
    \includegraphics[width=.9\linewidth]
    {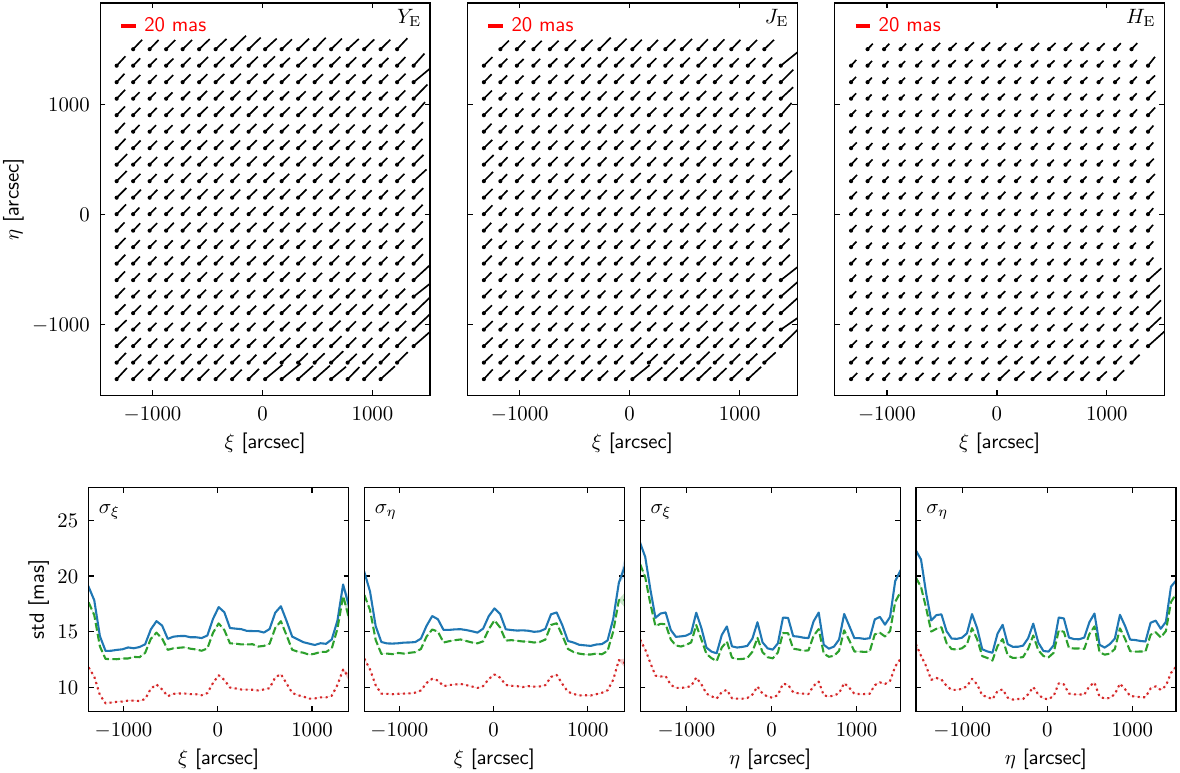}

\caption{As in Fig.~\ref{fig:fp_offsets}, but for the standard deviation of the
residuals. Whiskers have components $(\sigma_\xi,\sigma_\eta)$. The increased residuals at the edges results from the dithering pattern, which reduces the number of dithers in those areas.}
    \label{fig:fp_stdev}
\end{figure*}

\begin{figure}
    \centering
    \includegraphics[width=1\linewidth]{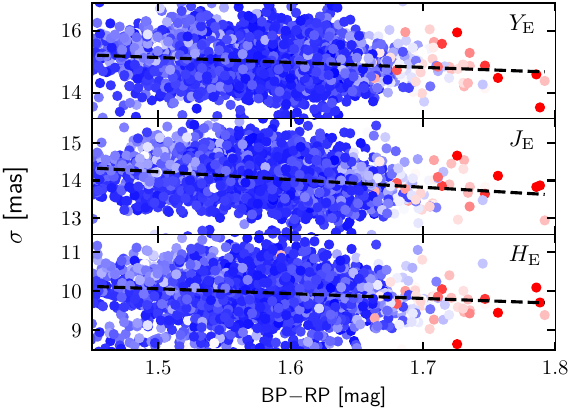}
   \caption{Standard deviation of the residuals in each stacked catalogue, summed in quadrature on $\alpha$ and $\delta$. This is displayed in relation to the average \Gaia colours of the matched sources. The Galactic extinction is represented by the colours: bluer hues indicate lower Galactic extinction, whereas redder hues correspond to higher values. The linear fit, depicted as a dashed line, has a slope of less than 3\,mas mag$^{-1}$ across all bands.}
    \label{fig:residuals_colors}
\end{figure}

\begin{figure*}
    \centering
    \includegraphics[width=.9\linewidth]{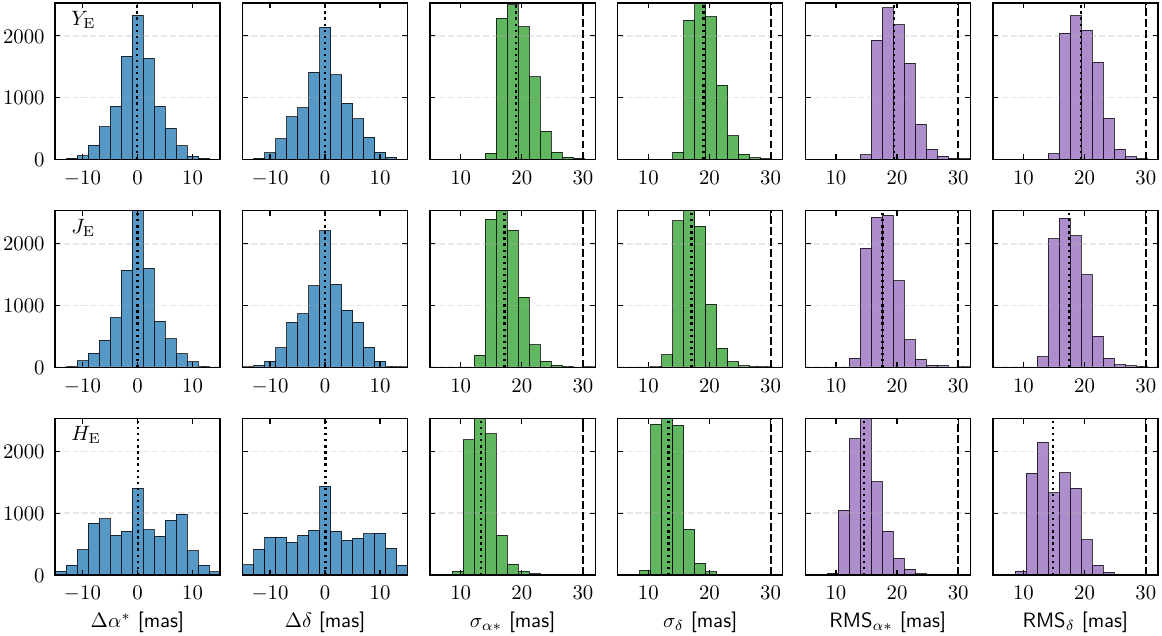}
    \caption{Distribution of astrometric residuals, including median offsets, standard deviations $\sigma$, and RMS, is displayed in the overlaps across the stacked catalogues. For the plots displaying $\sigma$ and RMS,
    the dashed line represents the legacy requirement (30 mas).
     Vertical dotted lines indicate the median values.}
    \label{fig:overlaps}
\end{figure*}

\subsection{Analysis of stacked catalogues\label{sc:analysis_stacks}}
To comprehensively assess the accuracy and robustness of the astrometric calibration of NIR images in the Euclid\ DR1, we present a series of diagnostic tests performed to investigate potential dependencies affecting the residuals. These tests examine variations in astrometric residuals as a function of Galactic extinction, source colour, observation time, and telescope rotation. Additionally, internal consistency checks analyse offsets between different filters and between the  NISP and VIS bands, as well as comparisons with the \Gaia catalogue. 
To accomplish this, we examine the stacked catalogues produced for each \OBSID rather than single dithers. 
Consequently, the results are independent of those derived from individual dithers during the astrometric calibration. A comparison of the residuals from the \Gaia \ positions obtained on dithers in \PE \ and those from the stacked catalogues reveals consistency, with a median standard deviation ($\sigma_\alpha$, $\sigma_\delta$) of approximately 10 mas, and 95\% of the dithers falling within 15 mas.

The centroids   of the sources  in the stacked catalogues, obtained by PSF-fitting \citep{Q1-TP003},  are cross-matched with the \Gaia catalogue by applying the  selection criteria described in Sect.~\ref{sc:Pipeline}. A strict mutual nearest-neighbour algorithm is adopted in the cross-match to minimise duplicate or incorrect assignments.  First, we perform a forward match, identifying the nearest \Gaia source for each target object in the stacked catalogue. Next, we perform a reverse match, identifying the nearest target object for every source in the \Gaia catalogue. A cross-match is considered if two criteria are met: (1) the target object and the \Gaia source are each other's closest reciprocal matches; and (2) their angular separation does not exceed a maximum tolerance  of 1\arcsec.

\subsubsection{Astrometric residuals versus position angle and time}

In the {\tt PV} formalism, distortion changes with the instrument's rotation relative to the sky.  
To test whether the position-angle-dependent distortion model introduces spurious residuals, we fitted, for each band, the RMS of the residuals per observation and their mean offsets in the instrument frame ($\Delta\xi$, $\Delta\eta$) with a sinusoidal function of the position angle. The amplitude of the sinusoid is 0.08–0.14 mas for the RMS, compared with a median RMS of approximately 10 mas per observation, and at most 0.08 mas for the mean offsets, compared with their uncertainty of 0.5–0.8 mas per observation. The position-angle dependence of the distortion model therefore introduces no significant spurious residuals.
Similarly, a linear fit of the RMS per observation as a function of time gives slopes of 0.04–0.11 mas yr$^{-1}$, i.e. a change of at most 0.2 mas over the period covered by the observations, which is again negligible.

\subsubsection{Astrometric residuals versus focal plane position}

To examine the discrepancies across the focal plane, we create an astrometric residual map for each filter.
For this purpose, we first collect  from all stacked catalogues  the detections matched with \Gaia. 
 When averaging frames with different rotations, offsets are diluted 
 if derived from sky coordinates. 
For this reason here we compute offsets from the reference pixel ($u$, $v$ in Eq.~\ref{eqn:SIP}) converted to arcseconds with the pixel scale ($\xi$, $\eta$), and derive the NISP$-$\Gaia residuals in the instrument frame  ($\Delta\xi$, $\Delta\eta$).
Since the scatter is nearly isotropic, the RMS is instead frame invariant  ($\sigma_\xi \sim \sigma_{\alpha*}$, $\sigma_\eta \sim \sigma_{\delta}$) .

The mean offsets and the scatter of the residuals $\Delta\xi$ and $\Delta\eta$ are
computed from the combined detections of all stacked catalogues, as a function of
the position $\xi$, $\eta$ in the stack frame.
For the two-dimensional maps (Figs.~\ref{fig:fp_offsets} and \ref{fig:fp_stdev}, top rows), the
$(\xi,\eta)$ plane is divided into non-overlapping cells of
$2.5\arcmin\times2.5\arcmin$, each containing on average $10^4$ detections.
In each cell, the mean and the standard deviation of $\Delta\xi$ and $\Delta\eta$
are computed after an iterative $3\sigma$ clipping of all the detections falling in
it; cells with fewer than 100 detections after clipping are discarded. Each cell is
represented by a whisker starting at the cell centre: its
components are $\langle\Delta\xi\rangle$ and $\langle\Delta\eta\rangle$ for the mean offsets,  and
$\sigma_\xi$ and $\sigma_\eta$ for the scatter.
To highlight trends along each axis, we also compute one-dimensional profiles
(Figs.~\ref{fig:fp_offsets} and \ref{fig:fp_stdev}, bottom rows) in independent strips $1\arcmin$ wide, running
along $\xi$ and $\eta$ across the whole field, each containing on average
$10^5$ detections; the same $3\sigma$-clipped statistics are applied. As a result of this analysis, across the focal plane offsets are below $\pm 5$ mas in all bands; the standard deviation in both $\xi$ and $\eta$ does not vary significantly and, consistently with previous results, decreases from \YE, \JE ( $\sim$ 15 mas) and \HE ($\sim$ 10 mas).

\subsubsection{Astrometric residuals versus source colour}
Figure~\ref{fig:residuals_colors} illustrates the standard deviations of the astrometric residuals measured in each stacked catalogue, as a function of the average \Gaia $\BPG-\RPG$ colours, with average redder colours corresponding to a higher Galactic extinction. A slight trend is noticeable, with a difference of less than 1\,mas between stacks with average bluer and redder colours. A  correlation between 
astrometric residuals and average source colours would be expected in the case of chromaticity effects in the PSF, where the measured centroid would be expected to vary with stellar colour and increase the astrometric scatter. A higher Galactic extinction shifts the selected \Gaia stars towards redder colours and narrows the colour distribution: in the case of a PSF chromaticity,  this would result in a smaller PSF variation across the sample than in fields with a low Galactic extinction, thus reducing the chromatic contribution to  the astrometric scatter. 
 A slight trend is noticeable, with a difference of less than 1\,mas between stacks at lower and higher extinction levels.
To further verify whether the astrometric calibration may be influenced by chromaticity effects, we examine the correlations between the residuals along the $x$ and $y$ axes of the detectors and \Gaia colours ($\BPG-\RPG$). For this purpose, \Gaia positions and those measured in the NISP stacks are first reprojected onto the focal plane, and the residuals along the detector axes ($\Delta \xi$, $\Delta \eta$) 
are computed and binned  in $\BPG-\RPG$ colours. 
The chromaticity terms, represented as $R_\xi$ and $R_\eta$, are defined as the slopes of these binned residuals against $\BPG-\RPG$: $R_\xi=\left<\Delta \xi\right>_{\rm{bin}}/\left({\rm{BP}}-{{\rm RP}}\right)_{{\rm bin}}$\,, $R_\eta=\left<\Delta \eta\right>_{\rm{bin}}/\left({\rm{BP}}-{\rm {RP}}\right)_{\rm{bin}}$ are finally derived. We find that for all filters, $R_\xi < 1\,{\rm mas\, mag^{-1}}$ and $R_\eta \ll 1\,{\rm mas\,  mag^{-1}}$.

\begin{figure}
    \centering
    \includegraphics[width=1\linewidth]{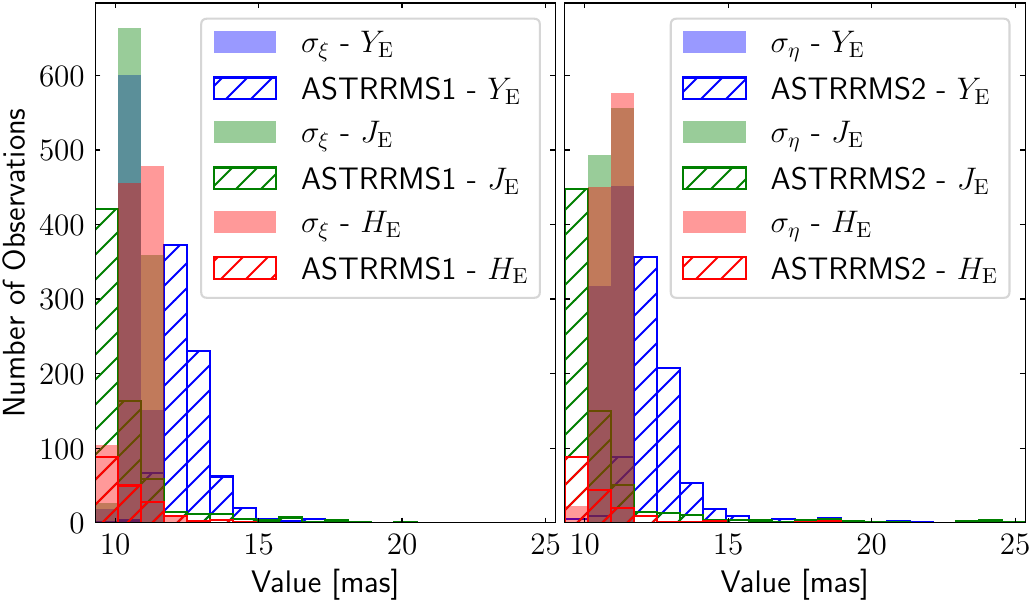}
   \caption{Comparison of the standard deviation of residuals when evaluating the impact of distortion calculated using \SCAMP versus a fixed distortion across the three NISP bands, in the self-calibration dithers. The hashed areas represent the standard deviations of residuals using the \Gaia catalogue as a reference, computed with \PE; the filled areas correspond to the values derived from the procedure outlined in the text, which assumes a constant distortion while varying only the linear terms.}
    \label{fig:distortion_consistency}
\end{figure}

\subsubsection{Astrometric consistency across overlaps
\label{sc:overlaps}}
Stacks of adjacent observations may overlap. This overlap enables us to assess whether the astrometric calibration remains consistent for the same sources that are detected in different stacks.
To accomplish this, we select pairwise overlaps between stacks,    with each having an overlap of at least  50\,arcmin$^2$, ensuring sufficient detections for a meaningful analysis. This results in a total area of 560\,$\deg^2$.
As shown in Fig.~\ref{fig:overlaps}, for 99\% of the overlaps the standard deviation in both coordinates remains within 25\,mas, while the median offsets are within 10\,mas. This is in agreement with what derived in the previous sections.

\subsection{Geometric distortion: accuracy and stability \label{sc:stability}}

We verify whether the  geometric distortion is calibrated  within 0.015\% over scales of 3\arcmin, as in the requirement. To achieve this, for each stacked catalogue we extract  within  the matched sources  the pairs separated  by a distance $d_{\rm ref}$ between \ang{;2.5;} and \ang{;3.5}  in the \Gaia catalogue, used as the reference.  We then compare these pairs with the corresponding ones in the \YE, \JE, and \HE catalogues, where the separation is $d_{\rm NIR}$, resulting in the fraction $\nu = |d_{\rm ref} - d_{\rm NIR}|/(\sqrt{2} d_{\rm ref})$\,.
A factor of $\sqrt{2}$ is introduced to transform 
pair separations into single-source equivalents. In all the stacks examined here, the RMS of $\nu$ is less than 0.01\%; in each stack, 95\% of the considered pairs have $\nu < 0.02\%$.
To ensure that this test accurately traces geometric distortions without being influenced by random centroiding noise, we conduct the analysis again on a 15\arcsec\ scale, where the impact of geometric distortion is minimal, and find the noise contribution to be negligible.

We then evaluate how well the assumption of a constant presolution aligns with the actual distortion observed in the dithers. For this purpose it is essential to separate the distortion, assumed to be constant, from the components that may vary across different observations: here we assume that they may be described by  offset and linear terms.
This is performed
using a set of artificial source positions, 
following the procedure outlined below.
\begin {enumerate}
\item A list of artificial sources positioned on a grid ($x$, $y$) that matches the detector's size of $2048 \times 2048$ pixels is created.
\item This list is transformed into intermediate world coordinates (IWC) using the {\tt PV} distortion computed by  \PE, yielding the reference coordinates ($\xi_{\rm r}$, $\eta_{\rm r}$).
\item The {\tt PV} distortion model contained in the presolution is converted to {\tt SIP}.
Equation~(\ref{eqn:SIP}) is applied to transform pixel coordinates \((x,y)\) to intermediate coordinates \((U,V)\), using only the nonlinear terms of the {\tt SIP} distortion ($i+j \geq 2$).
\item A least squares minimisation is performed to fit the $C_{ij}$ terms  of the {\tt CD} matrix (Eq.~\ref{eqn:PV}) and an offset \((\Delta \xi, \Delta \eta)\):
\begin{equation}
        \xi_{\rm r} = C_{11} U + C_{12} V + \Delta \xi \, ; \qquad \eta_{\rm r} = C_{21} U + C_{22} V + \Delta \eta\,.
        \label{eqn:lsq}
\end{equation}
\item The offset \((\Delta \xi, \Delta \eta)\) is converted back to pixel offsets \((\Delta U, \Delta V)\) by applying the inverse {\tt CD} matrix:
\begin{equation}
    \Delta U = \frac{C_{22} \Delta\xi - C_{12} \Delta\eta}{|C|} \, ; \qquad \Delta V = \frac{C_{11} \Delta\eta - C_{21} \Delta\xi}{|C|} \, .
\label{eqn:invtr}
\end{equation}
Here $|C| = C_{11}C_{22} - C_{12}C_{21}$\,.
The offsets $\Delta U$, $\Delta V$ replace the constant terms in the SIP polynomials, \(A_{0,0}\) and \(B_{0,0}\). Because the linear terms are now represented by the fitted {\tt CD} matrix, the linear {\tt SIP} terms \(A_{1,0}\), \(A_{0,1}\), \(B_{1,0}\), and \(B_{0,1}\) are set to zero to avoid redundancy.
\item The reference coordinates $\xi_{\rm r}$, $\eta_{\rm r}$ are compared with the coordinates $\xi_{\rm{n}}$, $\eta_{\rm{n}}$ obtained by applying the new distortion terms.
\end{enumerate}

We analyse 15 observations of the self-calibration field, each comprising 76 dithers with varying rotation angles. 
This selection is made because of the notably stable distortion observed in these cases. The ($\xi_{\rm{n}}$, $\eta_{\rm{n}}$) values obtained from the pixel grid are compared with those derived by applying the \SCAMP distortion contained in the headers ($\xi_{\rm{r}}$, $\eta_{\rm{r}}$), and the offsets are analysed. Figure~\ref{fig:distortion_consistency} illustrates the comparison between the standard deviation of these offsets and those computed by  \PE with respect to the \Gaia sources ({\tt ASTRRMS1}, {\tt ASTRRMS2}), based on the \SCAMP processing. In both methods, the median of the standard deviations is approximately 10\,mas, and within 20\,mas for 99\% of them.  
These findings suggest that the distortion remains consistently stable over time, within a margin of approximately 10\,mas. However, they also imply that the assumption that linear components can sufficiently account for the residual variations in distortion may not hold true, as further discussed in Sect.~\ref{sc:distvar}.

\begin{figure} 
\centering 
\includegraphics[width=1.0\linewidth]{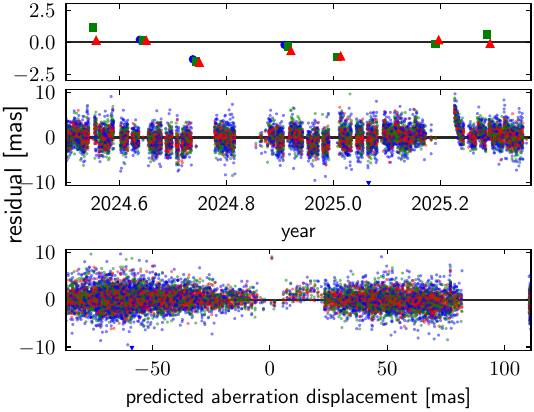} 
\caption{Residuals between the measured and predicted  aberration
plate-scale change, expressed as the RMS
displacement over the NISP field of view.
Top: self-calibration visits as a function of time; error bars
combine the error of the median over the dithers and the intrinsic
visit-to-visit scatter.
Middle: individual wide-survey fields as a function of time.
Bottom: fields used in the fit as a function of the predicted
aberration displacement.
Symbols are coloured by band (\YE: blue, \JE: green, \HE: red);  triangles at the panel edges
mark points beyond $\pm10$\,mas.}

\label{fig:aberration}
\end{figure}

\subsection{Effect of stellar aberration and plate-scale stability}
\label{sec:aberration}

The \Gaia reference positions used in the astrometric calibration are
barycentric and are propagated to the epoch of each exposure using proper
motions only; no correction for stellar aberration due to the motion of
\Euclid is applied. To first order, aberration changes the apparent angular
separations across the field by a factor $(1+\mathbf{v}\cdot\hat{\mathbf{n}}/\mathrm{c})$,
where $\mathbf{v}$ is the barycentric velocity of \Euclid, given by the \texttt{VEL\_X}, \texttt{VEL\_Y}, \texttt{VEL\_Z} values in the frame headers,  and $\hat{\mathbf{n}}$
the direction of the field, derived from the \texttt{CRVAL1} and \texttt{CRVAL2} values. Since the astrometric solution maps the observed
pixel positions onto barycentric reference positions, this term is absorbed
into the fitted plate scale $s$, which is then expected to vary by $\Delta s/s = \mathbf{v}\cdot\hat{\mathbf{n}}/c$ \citep[see e.g.][]{Cox_2003}.

For each detector $d$ in a dither we compute the effective pixel scale,
$s_d$, as the mean pixel size given by the full astrometric solution,
including the distortion terms. To remove the differences in scale between
detectors, caused e.g.\ by the optical distortion, we consider  
the fractional change $\Delta s_d/s_d$ relative to the median of $s_d$ over all
dithers in all available fields. The scale change of a dither is the median of
$\Delta s_d/s_d$ over the detectors, and the scale change, $\Delta s/s$, of a field is the
median over its dithers.
These values are then fitted as
\begin{equation}
  \frac{\Delta s}{s} = k\,\frac{\mathbf{v}\cdot\hat{\mathbf{n}}}{\mathrm{c}} + \mathrm{const}\,,
  \label{eq:aberr}
\end{equation}
where $k=1$ if the scale changes are reproduced by aberration, and the
constant accounts for the arbitrary zero point.
To express the results as displacements on the sky, we convert the  fractional scale change into the RMS displacement it produces over the
NISP field of view, $\delta = (\Delta s/s)\,r$, where
$r = 18\arcmin$  
is the RMS distance of the detector pixels from the centre of the field,
computed from the astrometric solution. Likewise, the predicted displacement is $\delta_\mathrm{pred} =r\, \left(\mathbf{v}\cdot\hat{\mathbf{n}}/c\right)$,.

Figure~\ref{fig:aberration} shows the residuals between the measured and predicted ($k=1$ in Eq.~\ref{eq:aberr}) scale change, as a function of time and of
the predicted displacement.  
We restrict the analysis to observations taken after July 2024, i.e. after the second thermal decontamination campaign \citep{DR1-TP026b}, when \Euclid had reached stable thermal conditions.
In the self-calibration field, the displacement of stars due to aberration is close to its maximum, $\sim$ 20\arcsec, but it is a nearly uniform shift absorbed by the pointing. This field lies close to the North Ecliptic Pole, so  the barycentric direction of its centre is almost perpendicular to 
the direction of motion of \Euclid relative to the Solar System barycentre. The effect on the scale is therefore small, and corresponds to an absolute displacement of approximately 3\,mas. Since the predicted displacement is small and the number of measurements lower, the constraint on $k$ is correspondingly weaker, but still consistent with $k=1$ within uncertainties. 
In the wide survey fields, instead, the predicted aberration displacement is within $\pm 100$\,mas. 
The fit gives $k = 0.997 \pm 0.001$,
consistent within the uncertainties in the three bands. The measured
scale changes thus follow the expected aberration; after removing it,
68\% of the fields deviate from the prediction by less than
2\,mas, 
with no significant mean offset. 

\subsection{Parallax}

As a further consistency check, we verify that the
positions of stars in the self-calibration field observed at different epochs contain the parallax displacement expected for each star from \Gaia.
Since the astrometric distortion is a  smooth function of position on the detector,
it can  absorb a displacement common to all stars on a detector, but
not the part that differs from star to star with their individual parallaxes.
That part should remain fully in the calibrated positions, with the amplitude
predicted from the \Gaia parallax and the position of
\Euclid at each epoch.
We test this with the observations of the
self-calibration field, by measuring the parallax of individual stars from the
calibrated positions alone and comparing it with \Gaia.
The test is limited at faint magnitudes by the precision of the
\Gaia parallaxes and at bright magnitudes by saturation in  NISP; since nearby red stars are  brighter in the near-infrared than
in the \GG band, the effect of these limits is to remove the reddest and nearest stars, and keep 
mostly red dwarfs with  parallax $\varpi_{\textit{Gaia}}\approx2$--$5$\,mas,
and few with larger parallaxes.
Moreover, the visits of the self-calibration
field span about 1 year, too short to separate parallax from proper motion,
which is therefore fixed to the \Gaia value. Since the
self-calibration dither pattern has large offsets, a star falls on the detectors in only a fraction  of the 76 dithers in a visit, from about 60\%
near the centre of the pattern to a few per cent at its edge. 

Positions in the calibrated NISP catalogs are  matched (within
$0\farcs3$) to \Gaia stars brighter than $\GG=20$, $\texttt{ruwe}<1.4$ and
$\texttt{astrometric\_excess\_noise}<1$\,mas \citep{Lindegren_21a} that are not flagged as
duplicated sources, and the residual
$\vec{r}=(\Delta\alpha^{*},\Delta\delta)$ with respect to the \Gaia
position propagated with proper motion only is computed at each exposure epoch. The barycentric position of \Euclid, $\vec{b}$, is taken as the
barycentric position of the Earth plus the geocentric position of the
spacecraft given in the \texttt{POS\_X/Y/Z} header keywords. With
$\vec{b}=(X,Y,Z)$ expressed in au in ICRS axes, the parallax factors of a star
at $(\alpha,\delta)$ are, following the convention of the \textit{Gaia}
astrometric model,
\begin{equation}
\begin{aligned}
F_{\alpha*} &= X\sin\alpha - Y\cos\alpha\, ;\\
F_\delta &= X\sin\delta\cos\alpha + Y\sin\delta\sin\alpha - Z\cos\delta\, ,
  \label{eq:plxfactor}
\end{aligned}  
\end{equation}
so that a star of parallax $\varpi$ is displaced by
$\varpi\vec{F}=\varpi\,(F_{\alpha*},F_\delta)$ in $(\alpha_*,\delta)$.

Two sets of stars are used. \emph{Probe} stars have a precise \Gaia
parallax ($\varpi>2$\,mas, $\varpi/\sigma_\varpi>10$).
\emph{Anchor} stars are all other \Gaia stars, regardless of their parallax.
For each exposure and detector, the mean offset of the calibrated positions
is computed from the anchors as
$\vec{c}=\langle\vec{r}-\varpi_{\textit{Gaia}}\vec{F}\rangle$ (with
$4\sigma$ clipping), i.e.\ after removing the parallactic displacement of each
anchor predicted from its own \textit{Gaia} parallax. Without this
subtraction, the mean anchor parallax ($\approx$ a few tenths of a mas) would
be absorbed by $\vec{c}$, biasing every probe parallax low by the same amount.
The anchor-corrected probe offsets $\vec{y}=\vec{r}-\vec{c}$ and the parallax
factors are  then averaged over the dithers of each visit.
  For each probe with
at least ten visits, $\varpi_{\rm NISP}$ is obtained by a least-squares fit
of $\vec{y}_i=\varpi_{\rm NISP}\vec{F}_i+\vec{o}$ to its visit means, where
the constant $\vec{o}$ absorbs the error of the propagated \textit{Gaia}
position, and its uncertainty $\sigma_{\rm NISP}$ from the scatter of the
star's own residuals about the fit.

Due to the significant uncertainty in individual parallaxes, the comparison relies on the entire probe sample. We fit
\begin{equation}
  \varpi_{\rm NISP} = k\,\varpi_{\textit{Gaia}} + \delta
  \label{eq:line}
\end{equation}
by least squares with weights
$w_j=1/(\sigma_{{\rm NISP},j}^2+\sigma_{\varpi,{\textit{Gaia}},j}^2)$.
A loss of part of the parallactic signal in the calibration would give
$k<1$, a zero-point offset would give $\delta\neq0$. 
 The difference between the NISP and \Gaia
parallaxes can be expressed \citep{Best_2020} in units of their
uncertainties,
$z=(\varpi_{\rm NISP}-\varpi_{\textit{Gaia}})/
(\sigma_{\rm NISP}^2+\sigma_{\varpi,\textit{Gaia}}^2)^{1/2}$.

\begin{figure}
  \centering
  \includegraphics[width=\columnwidth]{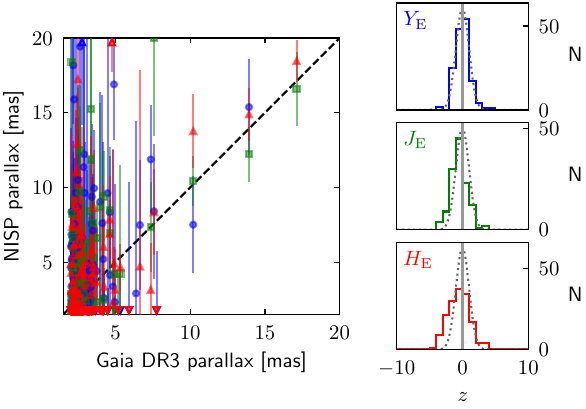}
  \caption{
  Parallaxes of the probe stars measured from the calibrated NISP
positions of the self-calibration field, versus their \textit{Gaia} DR3
parallaxes. Left: \YE (blue circles), \JE (green squares), and \HE
(red triangles), for the stars with at least ten
visits; error bars are the per-star uncertainties $\sigma_{\rm NISP}$,
derived from each star's own visit-to-visit scatter. Open triangles at
the top and bottom edges mark NISP parallaxes outside the plotted range.
The dashed line is the 1:1 relation. Right: distributions of the
normalised differences $z$ in \YE, \JE, \HE (top to bottom),
compared with a unit Gaussian scaled to the number of stars (dotted).}
  \label{fig:parallax}
\end{figure}

Figure~\ref{fig:parallax} compares the NISP and \Gaia parallaxes
for 152, 126, and 157 probe stars in \YE, \JE and \HE.
In \JE and \HE we find
$k=1.03\pm0.11$ and $1.12\pm0.16$: the full parallactic amplitude is
recovered. The same fits indicate a zero-point offset
$\delta\approx-1$\,mas in these bands ($-0.92\pm0.49$ and
$-1.18\pm0.52$\,mas), significant at about $2\sigma$ in each band. In \YE, the most undersampled band and hence the one with the lowest
astrometric precision, the constraints are weaker ($k=0.73\pm0.29$,
$\delta=+1.0\pm0.9$\,mas) and consistent both with $k=1$, $\delta=0$ and
with the \JE and \HE values. 
The median of $z$, with uncertainties bases on bootstrap resampling, is $+0.01\pm0.09$ (\YE), $-0.47\pm0.12$ (\JE), and $-0.43\pm0.18$ (\HE), consistently with with the slight offset found above. The robust RMS is 1.07, 1.31, and 1.67 respectively. The negative medians in $J$ and $H$
reflect the zero-point offset discussed below.

We conclude that the NISP astrometric calibration preserves the stellar
parallax: over the probe sample, the parallaxes recovered from the
calibrated positions have an amplitude consistent with that of
\Gaia, to 11--16\% in \JE and \HE, the two bands that constrain it
best, and at most $\approx20\%$ of the signal could have been lost
($2\sigma$). The possible zero-point offset of $\approx-1$\,mas is marginal and well within the astrometric accuracy.

\subsection{Inaccuracies due to persistence masking\label{sc:persistence} }

As anticipated in Sect.~\ref{sc:analysis_dithers}, 
the validation of the DR1 data products indicated that, for a specific number of dithers, the residuals in certain detectors, particularly DET13 and DET14, are higher than those in other detectors, exceeding  100\,mas. A comprehensive analysis revealed that, in these instances, the flagging of pixels is affected by persistence.
Persistence can cause false detections, and as detailed in \citet{DR1-TP003}, pixels in the persistence image that exceed a threshold of 2\,e$^-$ are considered invalid in the NISP images. However, this choice leads to a notable decrease in the number of sources with matches in the \Gaia catalogue, reducing from approximately 30 per detector to just a few, which in turn results in less accurate astrometric calibration. Shortly after this issue was identified, a solution was implemented (Sect. ~\ref{sc:persistence_corr}); however, it could not be incorporated into DR1.

To better estimate the impact of these failures on DR1, the following analysis has been conducted. For each stacked catalogue generated in DR1 for a specific \OBSID, detected sources are matched to the \Gaia catalogue within a 
5\arcsec\ radius to account for catastrophic astrometric failures.  
A similar procedure is employed to compare the NISP and VIS positions. In this case, VIS sources are initially selected with $|\texttt{spread\_model}| < 0.05$ in the stacked VIS catalogues \citep{Q1-TP002} to exclude overly extended sources. 

An RMS map is subsequently constructed utilising a 2D rolling window approach. The area covered by each stacked catalogue is divided into a grid with 30\arcsec\ intervals: this step size 
is chosen to achieve a finer spatial resolution and more detailed data sampling across the observed area. At the centre of each grid cell, matched sources are gathered within a specified radius, selected to ensure that each cell contains at least 10 (20) matched \Gaia (VIS) sources. This results in a radius of 4\arcmin\ for \Gaia and 2\arcmin\ for VIS; in the latter case, the radius is smaller because aligning with the VIS catalogue increases the number of matches with fainter NISP sources. If a cell contains fewer than the minimum required matches, the radius is incrementally increased by 10\arcsec\ until the requirement is met. On average, each cell contains 30, 50 matches with \Gaia   and VIS respectively.

\begin{figure}
    \centering
    \includegraphics[width=1\linewidth]{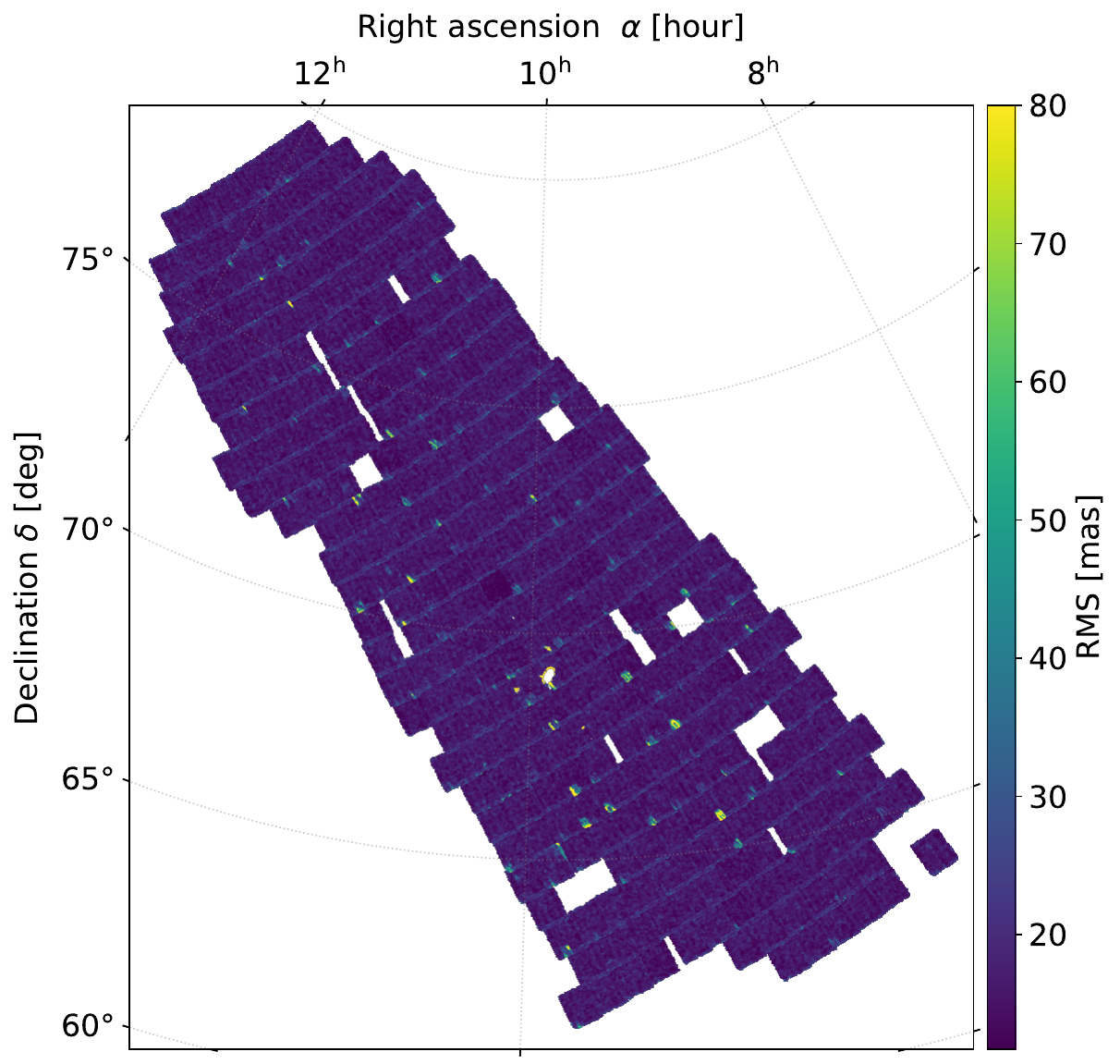}
        \includegraphics[width=1\linewidth]{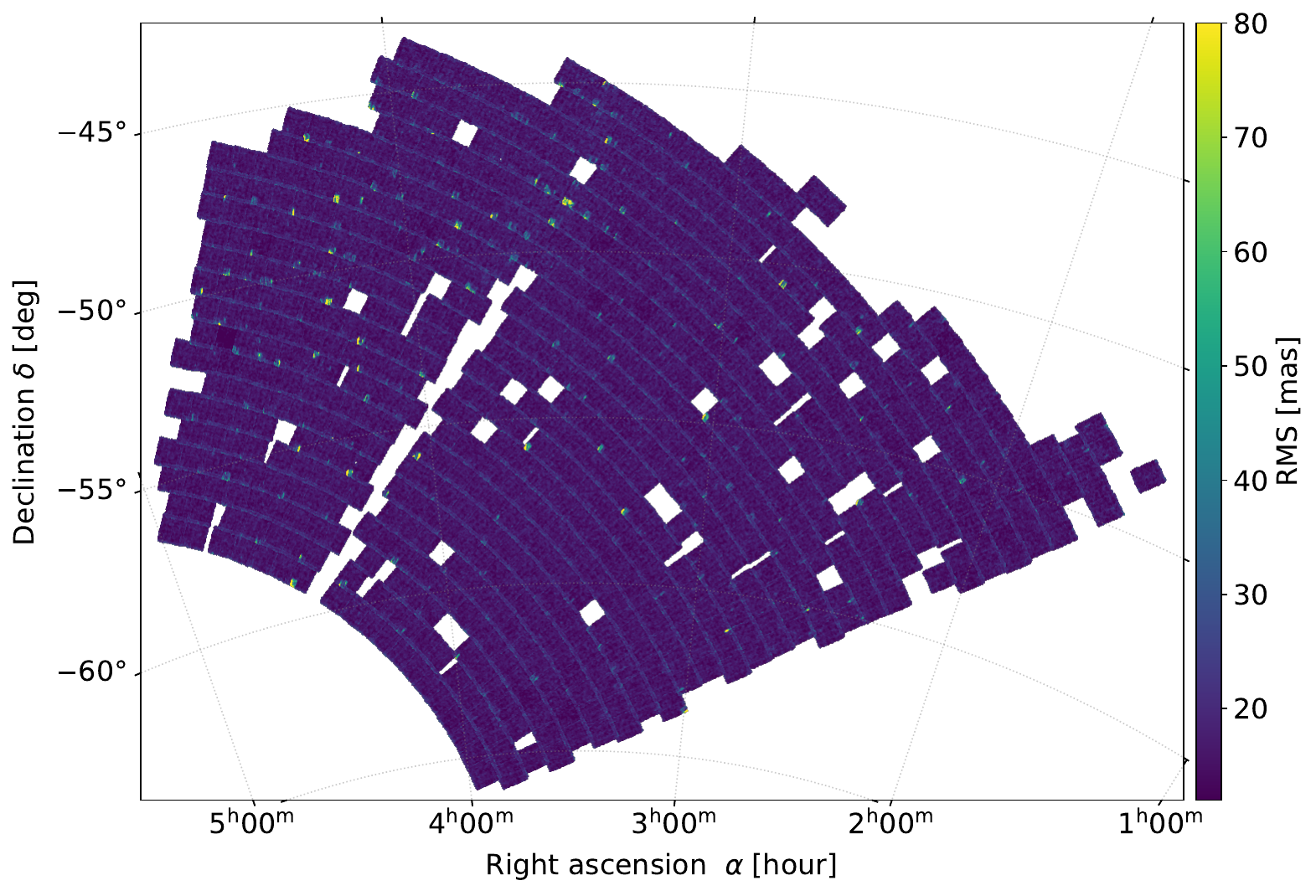}
   \caption{RMS maps derived from the spatial analysis of the \YE, \JE,  and\HE detections compared to VIS. The small green/yellow spots indicate areas where persistence flagging has led to reduced astrometric accuracy. }
    \label{fig:rms_maps}
\end{figure}

This approach enables us to identify regions where the RMS surpasses a predetermined threshold without significantly increasing the computational time. For each \OBSID,  a specific RMS map is generated. assigning the highest RMS values observed across  the NISP bands to each element of the map. The results of the comparison with VIS and \Gaia (Fig.~\ref{fig:rms_maps}) 
are consistent: the RMS exceeds 30, 60, and 90\,mas  in 3\%,  0.3\%, and 0.1\%  respectively of the total area, with a median RMS of approximately 20\,mas.
Finally, the same procedure is repeated using \HE as a reference and measuring the RMS in \JE and \YE. This produces results similar to those obtained previously, with RMS exceeding 30, 60, and 90 \,mas in 1\%, 0.2\%, and 0.1\% respectively of the area for each of \JE and \YE compared to \HE.

These tests indicate that despite this issue related to persistence masking, which will be resolved in Data Release 2 (Sect. ~\ref{sc:persistence_corr}), the astrometry remains consistent when compared to VIS and across the three NISP channels. The RMS is less than 30\,mas (0.1\,pixels) for at least 97\% of the surveyed area when comparing VIS to NISP. Across the NISP bands, the RMS  is less than   30\,mas (0.1 pixels) for 99\% of the area.

\begin{figure}
    \centering
    \includegraphics[width=1\linewidth]{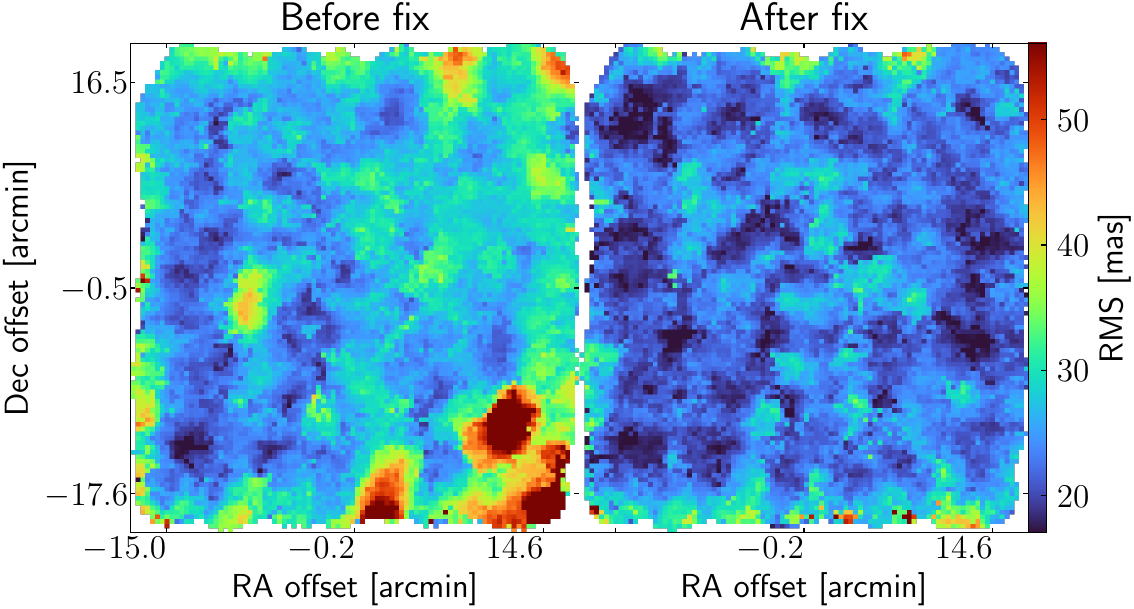}
   \caption{RMS maps  derived for a single observation ID both before and after applying the fix to mitigate the effects of persistence flagging. Initially, this led to a region with ${\rm RMS}>60\,{\rm mas}$  (left), which vanished after the fix was implemented (right).}
    \label{fig:rms_fix_comparison}
\end{figure}

\section{\label{sc:improvements}Improvements for DR2 and beyond}

\subsection{Reducing the impact on astrometry of the persistence masking\label{sc:persistence_corr}}

Although it was not feasible to rectify the persistence issue in DR1, a solution has been  developed and applied to subsequent processing. 
Astrometric calibration can be influenced by persistence, either through false detections that result in spurious matches or by false detections near real sources that disrupt the centroid estimate.
The impact of persistence flagging can be mitigated if a detection is discarded only when the pixels within a defined radius $R_{\rm{p}}$ from the detection centre exceed a specified persistence threshold $I_{\rm{p}}$. This ensures that only genuinely spurious detections are rejected, or those detections whose centre estimates may be compromised by persistence features. The values of 
$I_{\rm{p}} $ and $R_{\rm{p}}$ are determined following an analysis conducted on a subset of dithers: $I_{\rm{p}} = 5 \text{e}^-$ \citep[see also][]{DR1-TP038} and $R_{\rm{p}} = 5$ pixels. Figure~\ref{fig:rms_fix_comparison} demonstrates the impact of the fix on the astrometric residuals on a single observation ID.

\subsection{Exploring a simplified distortion model: preliminary Investigation\label{sc:presolution_linear} \label{sc:distvar}}
In DR1, the presolution is merely utilised in \SCAMP to enhance the initial matching of detected and \Gaia sources. As discussed in Sect.~\ref{sc:stability}, the distortion appears to be stable enough to consider keeping it constant while fitting only the linear terms. 

This alternative approach involves directly applying the presolution distortion model, assumed to be constant, to the source positions, followed by the computation of only the linear terms that are specific to each dither.
This simplified model would reduce computational complexity and minimise the sensitivity of the astrometric calibration to noise or spurious variations, potentially mitigating the risk of  overfitting.

\begin{figure}
    \centering
    \includegraphics[width=1\linewidth]{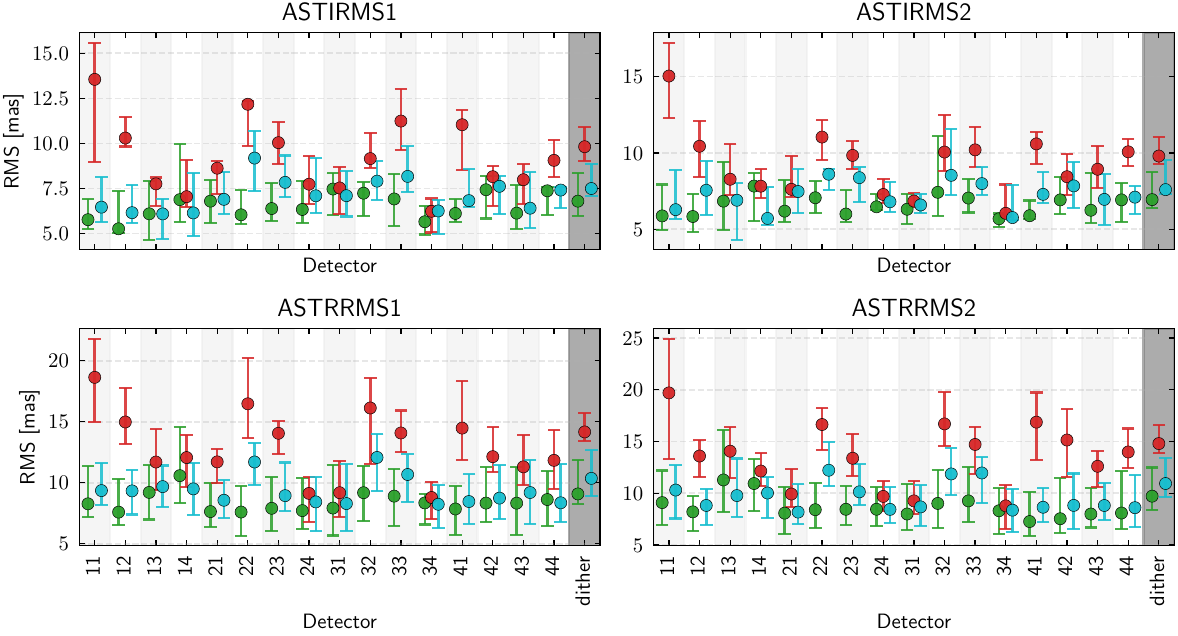}
   \caption{RMS residuals calculated for a subset of 10 observations (40 dithers), both for individual detectors and across the entire dither. The colours illustrate RMS values, including the median and 25\%/75\% percentiles, derived from various methods across detectors: green represents a comprehensive polynomial fit (current pipeline); red indicates a scenario with constant distortion where only linear terms are fitted; and cyan reflects constant distortion followed by a second-order fit. The RMS over the dithers is shown at the end and is highlighted in grey. }
    \label{fig:fixdist_dithers}
\end{figure}

The underlying assumption here is  that the linear corrections capture the residual variations due to the NISP optics (Sect.~\ref{sc:Pipeline}).
To initially evaluate the validity of this assumption, we developed a 
 modified version of \PE employing an approach 
similar to that outlined in Sect.~\ref{sc:stability}, and
conducted a pilot study based 
on a limited set of 10 observations (40 dithers):  initially, the presolution is converted into the {\tt SIP} formalism, which is then employed to remove distortion from the pixel coordinates in the \SExtractor catalogue. \SCAMP is subsequently executed on this modified catalogue. Two options are considered: (1) no additional distortion is computed in this second step, with only the linear terms being refitted, and (2) a polynomial of order 2 is fitted. 
Figure~\ref{fig:fixdist_dithers}, which illustrates the RMS of the residuals, supports the conclusions presented in Sect.~\ref{sc:stability}: relying solely on linear terms results in an increased RMS of the residuals. To achieve RMS values comparable to those obtained with the full fitting (10\,mas), it is necessary  to incorporate an additional second-order component.  A simplified distortion model  with fewer free parameters would still allow for a reduction in  the computational complexity; however, further comprehensive testing across a larger dataset is required before considering  potential integration into the pipeline.

\section {Conclusions}
This paper has presented the methodology and results of the astrometric calibration applied to the NISP data in Euclid DR1. The calibration pipeline employs the \SCAMP software to model and correct instrumental distortions on a per-dither basis, using a polynomial distortion representation within the {\tt PV} formalism. The \Gaia catalogue serves as the external astrometric reference, with source selections and quality controls implemented to ensure robust matching and calibration accuracy.

Analysis of  4000  observations covering 
1950~$\mathrm{deg}^2$ and over 1 year of observations demonstrates that the NIR astrometric calibration achieves an average precision of approximately 10 mas, with 99\% of the dithers exhibiting residuals below 20\,mas in each detector. 
This corresponds to less than 0.1 pixels, and is within the requirement of 30\,mas.
The average offsets relative to \Gaia are negligible (within 2 mas), confirming the absence of significant systematic biases. 

The analysis of the stacked catalogues confirms these values, allowing  further tests.
Spatial mapping of the residuals across the focal plane shows stable performance, with minor variations at the edges attributed to the dithering patterns.
The astrometric consistency among  the neighbouring dithers is confirmed by matching sources in overlapping areas across the stacks.
The chromaticity effects on astrometry are minimal, with slopes below 1 mas mag$^{-1}$, and no significant temporal or telescope rotation dependencies is observed.
The distortion pattern is stable over time within approximately 10 mas, supporting the potential for a constant distortion model supplemented by an additional correction to capture the residual variations. However, tests indicate that fitting only linear terms increases the residuals, and a second-order polynomial fit is necessary to maintain a calibration accuracy comparable to that of the full polynomial model.

A limitation identified in DR1 is the impact of persistence masking on astrometric accuracy, particularly affecting detectors DET13 and DET14. Excessive flagging of pixels due to persistence reduces the number of matched sources and may locally degrade the calibration precision. Upon comparison with the VIS detections, it is found that this issue  (${\rm RMS}>30\,{\rm mas}$) impacts less than 5\% of the surveyed area. This test also allows us to verify the astrometric consistency across the different NISP bands.
A mitigation strategy involving a refined persistence flagging criterion based on persistence intensity and radius around detections has been developed and applied in subsequent processing, effectively restoring astrometric accuracy in the affected regions.

For future data releases, improvements include a refined persistence masking approach and exploration of  alternative distortion evaluation methods that reduce the complexity of the distortion model while preserving the calibration quality. These enhancements aim to increase the calibration robustness and computational efficiency.

In conclusion, the \Euclid NISP astrometric calibration in DR1 demonstrates high precision and stability, with well-characterised residuals and identified pathways for further improvement. The results provide a solid foundation for subsequent data releases, ensuring accurate celestial source positioning, which is essential for the mission's scientific objectives.

\begin{acknowledgements}
\AckEC \AckDRone \AckDatalabs
This work has made use of data from the European Space Agency (ESA) mission
{\it Gaia} (\url{https://www.cosmos.esa.int/gaia}), processed by the {\it Gaia}
Data Processing and Analysis Consortium (DPAC,
\url{https://www.cosmos.esa.int/web/gaia/dpac/consortium}). Funding for the DPAC
has been provided by national institutions, in particular the institutions
participating in the {\it Gaia} Multilateral Agreement.
\end{acknowledgements}

\bibliographystyle{aa}
\bibliography{Euclid,myrefs} 

\begin{appendix}
\end{appendix}

\end{document}